\documentclass[preprint,12pt,authoryear]{elsarticle} % review

\usepackage[a4paper,margin=2.5cm]{geometry}
\usepackage{xcolor}
\usepackage{amsmath}
\usepackage{stmaryrd}
\usepackage{subfig}
\usepackage{multirow}
\usepackage{hyperref}
\usepackage{pstool}
\usepackage{psfrag}

\newcommand{\sgn}[1]{\mbox{sgn}(#1)}
\newcommand{\ignore}[1]{}
\newcommand{\new}[1]{#1}

\usepackage{amssymb}
\hypersetup{colorlinks, bookmarksopen,linkcolor = blue, citecolor = blue}

\journal{arxiv}

\begin{document}

\begin{frontmatter}

\title{Localization Analysis of an Energy-Based Fourth-Order Gradient
Plasticity Model}

\author[affiliation]{Ond\v{r}ej Roko\v{s}\corref{mycorrespondingauthor}}
\cortext[mycorrespondingauthor]{Corresponding author.}
\ead{ondrej.rokos@fsv.cvut.cz}

\author[affiliation]{Jan Zeman}
\ead{zemanj@cml.fsv.cvut.cz}

\author[affiliation]{Milan Jir\'{a}sek}
\ead{milan.jirasek@fsv.cvut.cz}

\address[affiliation]{Department of Mechanics, Faculty of Civil Engineering,
Czech Technical University in Prague, Th\'{a}kurova 7, 166 29 Prague 6, Czech
Republic.}

\begin{abstract}
The purpose of this paper is to provide analytical and numerical solutions of
the formation and evolution of the localized plastic zone in a uniaxially loaded
bar with variable cross-sectional area. An energy-based variational approach is
employed and the governing equations with appropriate physical boundary
conditions, jump conditions, and regularity conditions at evolving
elasto-plastic interface are derived for a fourth-order explicit gradient
plasticity model with linear isotropic softening. Four examples that differ by
regularity of the yield stress and stress distributions are presented. Results
for the load level, size of the plastic zone, distribution of plastic strain and
its spatial derivatives, plastic elongation, and energy balance are constructed
and compared to another, previously discussed non-variational gradient
formulation.
\end{abstract}


\end{frontmatter}

%	SECTION 1
%-----------------------------------------------------------------------------
%

\section{Introduction}
\label{Sect:1}
The presence of a softening branch of the stress-strain curve,
usually caused by initiation, propagation and coalescence of defects such as micro-cracks or
micro-voids, is a phenomenon typical of quasi-brittle materials. 
Softening often leads to localization of strain into narrow bands whose width
is related to an intrinsic
length dictated by the heterogeneities of the material
microstructure. Softening can be conveniently incorporated into damage models,
but can also be described by plasticity with a negative hardening modulus.
However, constitutive models within the classical continuum framework of simple materials
do not contain any length scale reflecting the typical size of
microstructural features. Therefore, the localization processes due to softening
are not described properly and mathematical models lead to ill-posed problems accompanied by localization of strain into subdomains of
zero volume and consequently to vanishing dissipation. Various enrichments
incorporating some information about the material heterogeneity have been
developed. They utilize, for example, additional kinematic variables, weighted
spatial averages, higher-order gradients, or rate-dependent terms; see e.g.\ the 
comparative studies and review
papers by \cite{VreBreGil95}, \cite{Jirasek98SS}, \cite{PeeGeeBorBre01}, \cite{JirRol02}, \cite{BazJir02}, \cite{JirRol09a} and \cite{JirasRolshoPart2}.
 These
techniques on the one hand preclude localization and loss of ellipticity of the
governing equations, but on the other hand significantly complicate the overall
analysis. For instance, in the case of higher-order gradient models, the
regularity conditions of internal variables at the evolving elasto-plastic
interface are not easy to characterize.

The present paper is devoted to the investigation of the Aifantis explicit
fourth-order gradient plasticity model, cf. e.g.~\cite{ZbibAifantis}
or~\cite{MuhlhausAifantis}, under conditions leading to non-uniform stress
fields. To keep the analysis transparent, we confine ourselves to
one-dimensional tension tests and perform our analysis in the framework of the so-called energetic solutions introduced in an abstract setting in~\cite{RateMie} and in the context of finite-strain plasticity in~\cite{MultPlastMie}, generalizing earlier variational formulations of
damage~\citep{Francfort:1993:SDE} and fracture~\citep{Francfort:1998:RBF}.
Building on this basis, we will derive the governing
equations and appropriate boundary and jump conditions. In
particular, the often questioned regularity conditions for internal variables at
the elasto-plastic interface will emerge naturally in a consistent and unified
way. Detailed solutions for four different test problems with various regularity
of the yield stress and stress distributions will be presented and compared to
results obtained from the standard non-variational gradient formulation
available in~\cite{SoftJirZemVond}. The present work can also be viewed as a
continuation and extension of results provided in our previous
paper~\citep{LocAnalJir}, where the second-order gradient plasticity model
was investigated. The main difference from our previous work is that now we
treat a more complicated model with higher-order regularity requirements on
internal variables and more intricate conditions at the
elasto-plastic interface. In addition, we approach the problem utilizing the
nowadays standard variational framework for rate-independent evolution.

This study is also closely related to several one-dimensional
studies into energy-based second-order gradient models of strain-softening
damage and plasticity. In particular, \cite{Pham:2011:IUS} performed a detailed
analysis of stability and bifurcation of localized and homogenous states,
obtained in the closed form, for a parameterized family of gradient damage
models. These results were later refined by~\cite{Pham:2013:ODM}, who studied
size effects and snap-back behaviour at the structural scale predicted by the
same group of damage models. The combination of damage and plasticity has been
the subject of recent contributions by \cite{DelPiero:2013:DCA}
and~\cite{Alessi:2014:GDMC}, with the emphasis on the competition between
brittle and ductile failure; the former work, \new{as well as~\cite{Milasinovic}}, also contain a validation against
experimental data. \new{Theoretical results of fracture and plasticity as $\Gamma$-limits of damage models within one-dimensional setting are described in~\cite{Flaviana}, and non-local damage or fracture analyses in bars under tension are discussed in~\cite{JiZe:IJSS:2015} and \cite{Lellis}; all these works also employ a variational formulation.} Our analysis extends these contributions by treating a model
regularized by the fourth derivative of internal variables and by
deriving the regularity of internal variables directly from energy-minimization arguments,
rather than enforcing them through additional boundary conditions at the
moving elastic-inelastic interfaces.

The energetic formulation for rate-independent processes comprises several steps
and relies on two principles. In the abstract
setting~\citep{Mielke:2006:ERIS}, the state of the system within a fixed
time horizon~$T$ is described in terms of a "non-dissipative"
field~$\boldsymbol{u}(t,\boldsymbol{x})$, $\boldsymbol{u}(t)\in\mathcal{U}$,
$\boldsymbol{x}\in\Omega$, where~$\Omega$ denotes the spatial domain, and
a~"dissipative" field~$\boldsymbol{z}(t,\boldsymbol{x})$, $\boldsymbol{z}(t)\in\mathcal{Z}$,
which specifies the irreversible processes at time~$t\in[0,T]$. The state of the
system is fully characterized by the state
variables~$\boldsymbol{q}(t,\boldsymbol{x})$, $\boldsymbol{q}(t)=(\boldsymbol{u}(t),\boldsymbol{z}(t))\in\mathcal{Q}=\mathcal{U}\times\mathcal{Z}$.
Typically, $\boldsymbol{u}$ is the displacement field and $\boldsymbol{z}$ is the field
of internal variables related to the inelastic phenomena, such as plastic strain or damage.
Further, we consider the total free (Helmholtz) energy of the
body~$\mathcal{E} : [0,T]\times\mathcal{Q}\rightarrow\mathbb{R}\cup\{+\infty\}$
together with the dissipation
distance~$\mathcal{D}:\mathcal{Z}\times\mathcal{Z}\rightarrow\mathbb{R}\cup\{+\infty\}$
which specifies the minimum amount of energy spent by the continuous transition
from state~$\boldsymbol{z}^{(1)}$ to state~$\boldsymbol{z}^{(2)}$. For
notational convenience, we will sometimes refer to the dissipation distance by
$\mathcal{D}(\boldsymbol{q}^{(1)},\boldsymbol{q}^{(2)})$ instead of
$\mathcal{D}(\boldsymbol{z}^{(1)},\boldsymbol{z}^{(2)})$. Then, the
process~$\boldsymbol{q}:[0,T]\rightarrow\mathcal{Q}$ is an energetic solution to the initial-value problem described by~$(\mathcal{E},\mathcal{D},\boldsymbol{q}_0)$ if it satisfies
\begin{enumerate}[(i)]
	\item {\bf Global stability:} for all~$t\in[0,T]$ and for all 	$\widehat{\boldsymbol{q}}\in\mathcal{Q}$
	\begin{equation}
	\mathcal{E}(t,\boldsymbol{q}(t))\leq\mathcal{E}(t,\widehat{\boldsymbol{q}})+\mathcal{D}(\boldsymbol{q}(t),\widehat{\boldsymbol{q}})
	\tag{S}
	\label{Sect:1:Eq:S}
	\end{equation}
	which ensures that the solution minimizes the sum~$\mathcal{E}+\mathcal{D}$,
	\item {\bf Energy equality:} for all $t\in[0,T]$
	\begin{equation}
	\mathcal{E}(t,\boldsymbol{q}(t))+\mathrm{Var}_{\mathcal{D}}(\boldsymbol{q};0,t)=\mathcal{E}(0,\boldsymbol{q}(0))+\int_0^t\mathcal{P}(s)\,\mbox{d}s
	\tag{E}
	\label{Sect:1:Eq:E}
	\end{equation}
	which expresses energy balance in terms of the internal energy, dissipated energy~$\mathrm{Var}_{\mathcal{D}}$, and time-integrated power of external forces~$\mathcal{P}$,
	\item {\bf Initial condition:}
	\begin{equation}
	\boldsymbol{q}(0)=\boldsymbol{q}_0
	\tag{I}
	\label{Sect:1:Eq:I}
	\end{equation}
\end{enumerate}
The dissipation along a process~$\boldsymbol{q}$ is expressed as
\begin{equation}
\mathrm{Var}_{\mathcal{D}}(\boldsymbol{q};0,t) = \sup \left\{ \sum_{i=1}^n
\mathcal{D}(\boldsymbol{q}(t_{i-1}),\boldsymbol{q}(t_{i})) \right\}
\label{Sect:1:Eq:1}
\end{equation}
where the supremum is taken over all~$n\in\mathbb{N}$ and all partitions of the time interval~$[0,t]$, $0=t_0<t_1<\dots<t_n=t$. Together, the two principles~\eqref{Sect:1:Eq:S} and~\eqref{Sect:1:Eq:E} along with initial condition~\eqref{Sect:1:Eq:I} naturally give rise to an
\newline\noindent
{\bf Incremental problem:} for $k=1,\ldots,N$
\begin{equation}
\boldsymbol{q}(t_k)\in\underset{\widehat{\boldsymbol{q}}\in\mathcal{Q}}{\mbox{Arg min}}\left[\mathcal{E}(t_k,\widehat{\boldsymbol{q}})+\mathcal{D}(\boldsymbol{q}(t_{k-1}),\widehat{\boldsymbol{q}})\right]
\tag{IP}
\label{Sect:1:Eq:IP}
\end{equation}
amenable to a numerical solution, in which each step is realized as a
minimization problem,
e.g.~\cite{Ortiz:1999:VFV,Carstensen:2002:NCP,Petryk:2003:IEM}. The main
conceptual difficulty with this incremental problem is that it represents a
global minimization, which is computationally cumbersome and physically
difficult to justify for non-convex energies. It is reasonable, however, to
assume that stable solutions to~\eqref{Sect:1:Eq:IP} are associated with local
minima; for comparative studies into evolution driven by local
and global energy minimization see
e.g.~\cite{Mielke:2011:DEM,Braides:2014:LM,Roubicek:2015:MDL}. On the other
hand, the variational approach offers many advantages, among which we highlight
that it provides a unified setting for the analysis, allows for
discontinuities in space, incorporates the governing laws with boundary
conditions and provides regularity conditions at the elasto-plastic interface.

For the uniaxial displacement-controlled tension test, we can further specify
all the quantities introduced above in more detail. Displacement~$u(t,x)$, where
we have used the light face letter since~$u$ is now a scalar field, as a
function of the spatial coordinate~$x\in\Omega\subset\mathbb{R}$, represents the
"non-dissipative" component; the total linearized strain is simply the spatial
derivative~$u'(t,x)=\partial u(t,x)/\partial x$. The "dissipative" variables,
describing the irreversible processes, are plastic strain~$\varepsilon_p(t,x)$
and cumulative plastic strain~$\kappa(t,x)$,
i.e.~$\boldsymbol{z}(t,x)=(\varepsilon_p(t,x),\kappa(t,x))$.\footnote{Strictly speaking, the "non-dissipative" component should have read $u_{el}(x,t)=u(x,t)-\int\varepsilon_p(x,t)\,\mbox{d}x$, where~$\int\bullet\,\mbox{d}x$ denotes a primitive integral of a function~$\bullet$. Since such an affine transformation does not affect the solution, we adopt~$u$ instead of~$u_{el}$ as our primal variable in further considerations for convenience.} The corresponding function spaces are as follows:
\begin{subequations}
\label{Sect:1:Eq:2}
\begin{align}
u\in V_u(t)&=\left\{\widehat{u}\in W^{1,2}(\Omega)\,|\,\widehat{u}=u_D(t)\mbox{ on }\partial\Omega\mbox{ in the sense of traces}\right\}\label{Sect:1:Eq:2a}\\
\varepsilon_p\in V_{\varepsilon_p}&= W^{2,2}(\Omega)\label{Sect:1:Eq:2b}\\
\kappa\in V_\kappa&=\left\{\widehat{\kappa}\in W^{2,2}(\Omega)\,|\,\widehat{\kappa}\geq 0\right\}\label{Sect:1:Eq:2c}
\end{align}
\end{subequations}
where~$u_D(t)$ denotes prescribed displacements on the boundary (specifying the
Dirichlet boundary condition) and~$W^{k,2}$ stands for the space of all Lebesgue square-integrable functions with square-integrable generalized derivatives up to
order~$k$; later on, we will employ a subset~$W^{k,2}_0$ consisting of functions vanishing at the boundary, for details we
refer to e.g.~\cite{EvansPDE}. Consequently, we identify~$\mathcal{U}=V_u(t)$,
which now depends on time (due to the time-dependent values presented on the boundary),
$\mathcal{Z}=V_{\varepsilon_p}\times V_\kappa$,
$\mathcal{Q}=\mathcal{U}\times\mathcal{Z}=V_u(t)\times V_{\varepsilon_p}\times
V_\kappa$, and define the total free energy of the body
\begin{equation}
\mathcal{E}(t,u,\varepsilon_p,\kappa)=\int_\Omega\frac{1}{2}EA(u'-\varepsilon_p)^2\,\mbox{d}x+\int_\Omega\frac{1}{2}HA(\kappa^2-l^4\kappa''^2)\,\mbox{d}x-\int_\Omega Abu\,\mbox{d}x
\label{Sect:1:Eq:3}
\end{equation}
and the dissipation distance
\begin{equation}
\mathcal{D}(\boldsymbol{z}^{(1)},\boldsymbol{z}^{(2)})=
\left\{
\begin{aligned}
&\int_\Omega
A\sigma_0|\varepsilon_p^{(2)}-\varepsilon_p^{(1)}|\,\mbox{d}x&&\mbox{ if
}\kappa^{(2)}=\kappa^{(1)}+|\varepsilon_p^{(2)}-\varepsilon_p^{(1)}|\mbox{
in } \Omega\\ &+\infty && \mbox{
otherwise}
\end{aligned}
\right.
\label{Sect:1:Eq:4}
\end{equation}
Quantities appearing in the definitions of energies represent the Young
modulus~$E$~[Pa], softening modulus~$H<0$~[Pa], characteristic length of the
material~$l$~[m], initial yield stress~$\sigma_0$~[Pa], a function describing
the distribution of the cross-sectional area along the bar~$A(x)$~[m$^2$], and
prescribed body force density~$b(x)$~[N/m$^3$]. For completeness, let us note
that the power of external forces in~\eqref{Sect:1:Eq:E} has the
form~$\displaystyle{\mathcal{P}(s)=F(s)\dot{u}_D(s)}$, where~$F(s)$ denotes the
reaction force as a function of time and the dot stands for the time
derivative.

The paper is organized as follows. In Section~\ref{Sect:2}, we will
revisit the energetic formulation for the case of monotone loading, i.e.~$\dot{u}_D(t)\geq
0$, which greatly simplifies the specific form of~\eqref{Sect:1:Eq:S},
\eqref{Sect:1:Eq:E}, \eqref{Sect:1:Eq:I}, and~\eqref{Sect:1:Eq:IP}
accompanied by~\eqref{Sect:1:Eq:2}--\eqref{Sect:1:Eq:4}. Further, the governing equations
with boundary conditions, jump conditions, and regularity conditions at
the elasto-plastic interface will be derived for the resulting one-dimensional fourth-order
gradient-enriched plasticity model. Sections~\ref{Sect:3} and~\ref{Sect:4} are
concerned with piecewise constant yield stress and stress distributions, which
also represent the only cases amenable to analytical solutions. There, it will
be shown that the structural response may, in a certain range, exhibit hardening
due to the gradient enrichment despite the softening character of the material
model. In Sections~\ref{Sect:5} and~\ref{Sect:6}, the results for piecewise
linear and quadratic stress field distributions will be compared to standard
non-variational solutions available in~\cite{SoftJirZemVond}. In spite of all
the simplifications we will be forced to use numerical solutions. The
influence of data variation to the evolution of the plastic zone, its profile
and load-displacement diagrams will also be investigated. Finally, in
\ref{Sect:A}, the stability conditions for the case of a uniform bar are
discussed, and in \ref{Sect:B}, we verify optimality of the obtained regularity conditions at the elasto-plastic interface by an independent argument.
%
%-----------------------------------------------------------------------------
%	SECTION 2
%-----------------------------------------------------------------------------
%
\section{Energy-Based Formulation}
\label{Sect:2}
%
%----------------------------------
%	SUBSECTION 2.1
%----------------------------------
%
\subsection{General Considerations}
\label{SubSect:2.1}
For~$\mathcal{E}$ and~$\mathcal{D}$ given by~\eqref{Sect:1:Eq:3} and~\eqref{Sect:1:Eq:4}, minimization in~\eqref{Sect:1:Eq:IP} with respect to~$\widehat{\kappa}$ gives~$\kappa(t_k)=\kappa(t_{k-1})+|\widehat{\varepsilon}_p-\varepsilon_p(t_{k-1})|$.
Consequently, \eqref{Sect:1:Eq:IP} reduces to
\begin{equation}
\begin{aligned}
(u(t_k),\varepsilon_p(t_k))\in
\underset{(\widehat{u},\widehat{\varepsilon}_p)\in V_u(t_k)\times V_{\varepsilon_p}}{\mbox{Arg min}}
&\int_\Omega\frac{1}{2}EA(\widehat{u}'-\widehat{\varepsilon}_p)^2\,\mbox{d}x\\
&+\int_\Omega\frac{1}{2}HA[\kappa(t_{k-1})+|\widehat{\varepsilon}_p-\varepsilon_p(t_{k-1})|]^2\,\mbox{d}x\\
&-\int_\Omega\frac{1}{2}HAl^4[\kappa''(t_{k-1})+|\widehat{\varepsilon}_p-\varepsilon_p(t_{k-1})|'']^2\,\mbox{d}x\\
&+\int_\Omega A\sigma_0|\widehat{\varepsilon}_p-\varepsilon_p(t_{k-1})|\,\mbox{d}x-\int_\Omega Ab\widehat{u}\,\mbox{d}x
\end{aligned}
\label{SubSect:2.1:Eq:1}
\end{equation}
cf.\ also~Section~4.3 in \cite{MultPlastMie}, where the local plasticity model with
hardening is discussed. In Eq.~\eqref{SubSect:2.1:Eq:1}, spatial derivatives are
understood in the sense of distributions. For further considerations, we will
restrict ourselves to tensile loading with possible elastic unloading, but never with a reversal of the plastic flow. Then, the plastic strain~$\varepsilon_p$ and the cumulative plastic strain~$\kappa$ are equal, and
we can use~$\kappa$ as the only internal variable. As a result, instead of the
incremental approach given in Eqs.~\eqref{Sect:1:Eq:IP}
and~\eqref{SubSect:2.1:Eq:1}, it is fully sufficient to consider a total
formulation providing a parameterized solution~$(u(t),\kappa(t))$ which does not
violate the irreversibility constraints. Note that such a parametrization is
mathematically justified only when the elastic energy~$\mathcal{E}$ is strictly
convex in~$\boldsymbol{q}$, which implies that the solution is time-continuous
for sufficiently regular loading, cf.~\cite{RateMie}. \new{Later on, instead of imposing the Dirichlet boundary conditions, we prescribe directly the size of the plastic zone as a function of time in order to control the system evolution. This approach automatically entails time-continuity of~$\boldsymbol{q}(t)$, which implies satisfaction of the energy balance~\citep{Pham:2011:IUS,Pham:2013:ODM}, and also justifies the total formulation.}

Taking into account all the above simplifications, the minimization problem in Eq.~\eqref{SubSect:2.1:Eq:1} reduces to
\begin{equation}
(u(t),\kappa(t))\in
\underset{(\widehat{u},\widehat{\kappa})\in V_u(t)\times V_{\kappa}}{\mbox{Arg min}}\Pi(\widehat{u},\widehat{\kappa})
\label{SubSect:2.1:Eq:2}
\end{equation}
where
\begin{equation}
\begin{aligned}
\Pi(\widehat{u},\widehat{\kappa})=&\int_\Omega\frac{1}{2}EA(\widehat{u}'-\widehat{\kappa})^2\,\mbox{d}x+\int_\Omega\frac{1}{2}HA(\widehat{\kappa}^2-l^4\widehat{\kappa}''^2)\,\mbox{d}x\\
&+\int_\Omega A\sigma_0\widehat{\kappa}\,\mbox{d}x-\int_\Omega Ab\widehat{u}\,\mbox{d}x
\end{aligned}
\label{SubSect:2.1:Eq:3}
\end{equation}
which resembles a variational inequality of the first kind,
due to the requirement~$\kappa\geq 0$ in Eq.~\eqref{Sect:1:Eq:2c}.

The first variation (G\^{a}teaux derivative) of~$\Pi$ furnishes us with the optimality condition
\begin{equation}
\begin{aligned}
\delta\Pi(u,\kappa;\delta u,\delta\kappa)=&\int_\Omega EA(u'-\kappa)(\delta u'-\delta\kappa)\,\mbox{d}x+\int_\Omega HA(\kappa\delta\kappa-l^4\kappa''\delta\kappa'')\,\mbox{d}x\\
&+\int_\Omega A\sigma_0\delta\kappa\,\mbox{d}x-\int_\Omega Ab\delta u\,\mbox{d}x\geq 0
\end{aligned}
\label{SubSect:2.1:Eq:4}
\end{equation}
where~$\delta u$ and~$\delta\kappa$ are admissible variations satisfying~$(u+\delta u,\kappa+\delta\kappa)\in V_u(t)\times V_\kappa$ (for simplicity, explicit dependence on time~$t$ has been dropped). Employing the integration by parts we arrive at
\begin{equation}
\begin{aligned}
\delta\Pi(u,\kappa;\delta u,\delta\kappa)=&-\int_\Omega[(EA(u'-\kappa))'+Ab]\delta u\,\mbox{d}x\\
&+\int_\Omega[HA\kappa-(HAl^4\kappa'')''+A\sigma_0-EA(u'-\kappa)]\delta\kappa\,\mbox{d}x\\
&+\sum\limits_{\partial\Omega}EA(u'-\kappa)n\delta u-\sum\limits_i\llbracket EA(u'-\kappa)\delta u\rrbracket_{x_i}\\
&-\sum\limits_{\partial\Omega}HAl^4\kappa''n\delta\kappa'+\sum\limits_i\llbracket HAl^4\kappa''\delta\kappa'\rrbracket_{x_i}\\
&+\sum\limits_{\partial\Omega}(HAl^4\kappa'')'n\delta\kappa-\sum\limits_i\llbracket (HAl^4\kappa'')'\delta\kappa\rrbracket_{x_i}
\end{aligned}
\label{SubSect:2.1:Eq:5}
\end{equation}
where~$\sum_{\partial\Omega}$ denotes the boundary integral, in our one-dimensional setting reduced to the sum over two end points of the interval~$\Omega$, and~$n$ is the unit outer normal, which equals~$-1$ at the left~$\partial\Omega_L$ and~$1$ at the right~$\partial\Omega_R$ part of the boundary~$\partial\Omega=\partial\Omega_L\cup\partial\Omega_R$. The sums $\sum_i$ are taken over all points of possible discontinuity~$x_i$ and
\begin{equation}
\llbracket f\rrbracket_{x_i}=f(x_i^+)-f(x_i^-)=\lim_{x\downarrow x_i}f(x)-\lim_{x\uparrow x_i}f(x)
\label{SubSect:2.1:Eq:6}
\end{equation}
represents the jump of function~$f(x)$ at~$x_i$. The necessary condition for a local minimum is non-negativity of the first variation of the functional~$\Pi$ for all admissible variations~$\delta u$ and~$\delta\kappa$, cf. Section~8.4.2 in \cite{EvansPDE}, or Chapter~5 in \cite{RoubicekNonlin}. Below we show that such an approach leads to a consistent set of governing equations, namely the equilibrium equations, complementarity conditions of the plastic flow, boundary conditions, and regularity conditions at the elasto-plastic interfaces. Analysis of the second variation (second-order G\^{a}teaux derivative), which is related to stability of the solution~\eqref{Sect:1:Eq:S}, is postponed to~\ref{Sect:A}.
%
%----------------------------------
%	SUBSECTION 2.2
%----------------------------------
%
\subsection{Governing Equations}
\label{SubSect:2.2}
Since~$u+\delta u\in V_u$, we have~$\delta u\in W^{1,2}_0(\Omega)$, meaning
that~$\delta u$ is arbitrary inside~$\Omega$ with zero trace on the physical
boundary~$\partial\Omega$. Thus the expression multiplying~$\delta u$ in the
first line of~\eqref{SubSect:2.1:Eq:5} must vanish, providing us with the
static equilibrium condition
\begin{equation}
(EA(u'-\kappa))'+Ab=0\mbox{ in }\Omega
\label{SubSect:2.2:Eq:1}
\end{equation}
Here, $u'$ corresponds to the total strain, $u'-\kappa$ is the elastic strain, and~$EA(u'-\kappa)$ is the axial force, which is required to be continuous according to the third line of~\eqref{SubSect:2.1:Eq:5}, because~$\delta u$ is arbitrary inside~$\Omega$. Due to the zero trace of~$\delta u$, the sum over boundary points in the third line of~\eqref{SubSect:2.1:Eq:5} vanishes.

Since~$\kappa+\delta\kappa\in V_\kappa$, variations~$\delta\kappa$ cannot be
completely arbitrary. Let us define the plastic zone as the open set
$\mathcal{I}_p=\{x\in\Omega\,|\,\kappa(x)>0\}$, i.e., as the support of~$\kappa$, 
and the elastic zone as the open set~$\mathcal{I}_e=\{x\in\Omega\,|\,\kappa(x)=0\}$.
In $\mathcal{I}_p$, the variation $\delta\kappa$ can have an arbitrary sign,  
and so the expression multiplying~$\delta\kappa$ in the second line of~\eqref{SubSect:2.1:Eq:5} 
must vanish. On the other hand, only non-negative variations $\delta\kappa$
are admissible in $\mathcal{I}_e$,
and so the expression multiplying~$\delta\kappa$ does not necessarily vanish but
is constrained to be non-negative. The resulting conditions
\begin{eqnarray}
HA\kappa-(HAl^4\kappa'')''+A\sigma_0&=&EA(u'-\kappa)\mbox{ in }\mathcal{I}_p
\label{SubSect:2.2:Eq:2}
\\
HA\kappa-(HAl^4\kappa'')''+A\sigma_0&\geq& EA(u'-\kappa)\mbox{ in }\mathcal{I}_e
\label{SubSect:2.2:Eq:3}
\end{eqnarray}
combined with the definitions of $\mathcal{I}_p$ and $\mathcal{I}_e$ can be presented
in the complementarity format
\begin{eqnarray}\label{SubSect:2.2:Eq:3a}
\kappa &\geq& 0
\\
HA\kappa-(HAl^4\kappa'')''+A\sigma_0 -EA(u'-\kappa)&\geq& 0
\\
\left[HA\kappa-(HAl^4\kappa'')''+A\sigma_0- EA(u'-\kappa)\right]\cdot\kappa &=& 0
\label{SubSect:2.2:Eq:3c}
\end{eqnarray}
Note also that since $\kappa=0$ in $\mathcal{I}_e$, condition \eqref{SubSect:2.2:Eq:3}
could be simplified to
\begin{equation}
A\sigma_0\geq EAu'\mbox{ in }\mathcal{I}_e
\label{SubSect:2.2:Eq:4}
\end{equation}

For $AH=$~const., the second term on the left-hand side of Eq.~\eqref{SubSect:2.2:Eq:2} 
reduces to $-HAl^4\kappa^\mathrm{IV}$, and the standard formulation of the fourth-order
gradient plasticity model is recovered, 
cf.~\cite{SoftJirZemVond} and Tab.~\ref{SubSect:2.2:Tab:1}.
For variable sectional area and/or variable plastic modulus, 
expansion of the second term on the left-hand side of Eq.~\eqref{SubSect:2.2:Eq:2} 
gives~$-l^4[H(x)A(x)\kappa^\mathrm{IV}(x)+2(H(x)A(x))'\kappa'''(x)+(H(x)A(x))''\kappa''(x)]$.
With increasing magnitude of
the derivatives of~$H(x)A(x)$ we expect also increasing differences between the
solutions corresponding to the classical and variational formulations. 

In addition to conditions \eqref{SubSect:2.2:Eq:1}--\eqref{SubSect:2.2:Eq:3},
which have been deduced as optimality conditions following from the first two lines 
of~\eqref{SubSect:2.1:Eq:5},
the
last two lines of~\eqref{SubSect:2.1:Eq:5} provide us with boundary and
regularity conditions for the plastic strain. 
%at the elasto-plastic interface 
\ignore
{
Since the displacements are assumed to be prescribed on the whole boundary, their variations on the
boundary are zero and the first term in the third line of~\eqref{SubSect:2.1:Eq:5}
automatically vanishes. The second term implies continuity of $EA(u'-\kappa)$, i.e.,
of the normal force transmitted by the bar. A more refined discussion is needed
for the last two lines of~\eqref{SubSect:2.1:Eq:5}, which contain the variations
of plastic strain.  
}

Let us first discuss the {\bf boundary conditions}. Again, we have to distinguish between
the plastic part of the physical boundary, $\partial\Omega\cap\mathcal{I}_p$,
and the elastic part, $\partial\Omega\cap\mathcal{I}_e$. 
\begin{enumerate}
\item {\bf Boundary point in a plastic state}, characterized by $\kappa>0$:\\
The variation $\delta\kappa$ as well as its derivative $\delta\kappa'$ at such a point
can have an arbitrary sign and the terms that multiply them must vanish. This leads
to boundary conditions $\kappa''=0$ and $(HAl^4\kappa'')'=0$.
\item {\bf Boundary point in an elastic state}, characterized by $\kappa=0$:\\
The variation $\delta\kappa$ at such a point can only be zero or positive, and so the
term that multiplies $\delta\kappa$ in the first sum in the fifth line 
of~\eqref{SubSect:2.1:Eq:5}  must not be negative but does not need to vanish. 
Therefore, for a boundary point in an elastic state we obtain the inequality condition
$(HAl^4\kappa'')'n\ge 0$. Recall that $n=-1$ at the left boundary and $n=1$ at the right boundary.
Regarding the second condition, tested by the derivative of the variation of plastic strain,
we have to distinguish the following two subcases:
\begin{enumerate}
\item Nonzero derivative of plastic strain at the boundary:\\
If $\kappa=0$ and $\kappa'\ne 0$ at a boundary point (which necessarily means
$\kappa'>0$ at the left boundary and $\kappa'<0$ at the right boundary, by virtue
of the universally valid admissibility condition $\kappa(x)\ge 0$), then the variation of plastic strain, 
$\delta\kappa'$, has an arbitrary sign, and the term  $HAl^4\kappa''n$ that multiplies 
$\delta\kappa'$ in the first sum in the fourth line~\eqref{SubSect:2.1:Eq:5} must vanish.
Since $HAl^4\ne 0$, this gives the boundary condition $\kappa''=0$.
\item Zero derivative of plastic strain at the boundary:\\
If $\kappa=0$ and $\kappa'=0$ at a boundary point,
then the derivative of the variation of plastic strain, $\delta\kappa'$, can 
be positive at the left boundary and negative at the right boundary, which means that
  $\delta\kappa'n$ can be negative. Consequently, the term $HAl^4\kappa''$ that multiplies 
$\delta\kappa'n$ in the first sum in the fourth line~\eqref{SubSect:2.1:Eq:5} must not be
negative (note the minus sign before the sum). Since $H<0$ and $l^4A>0$, 
the resulting  condition reads $\kappa''\le 0$. But this actually cannot be satisfied 
as a strict inequality, because $\kappa''<0$ in combination with $\kappa=0$ and $\kappa'=0$
would lead to a violation of the admissibility condition $\kappa(x)\ge 0$ in the near vicinity
of the boundary point. So once again, we conclude that $\kappa''$ must vanish.
\end{enumerate} 
\end{enumerate} 
We have found that the boundary condition $\kappa''=0$ applies independently
of the state of the material at the boundary. On top of that, we have
  $(HAl^4\kappa'')'=0$ and $\kappa>0$ if the boundary point is in a plastic state,
and $(HAl^4\kappa'')'n\ge 0$ and $\kappa=0$ if the boundary point is in an elastic state.
All this can be summarized by the following boundary conditions:
\begin{eqnarray}
\kappa''&=&0 \\
(HAl^4\kappa'')'n&\ge& 0 \\
\kappa &\ge & 0 \\
(HAl^4\kappa'')'n\kappa &=& 0
\end{eqnarray}

\ignore{ ============= OLD STUFF ======================
as can be deduced
from the fourth and fifth line of~\eqref{SubSect:2.1:Eq:5} since~$\delta\kappa$
and~$\delta\kappa'$ are arbitrary. On the elastic
part~$\partial\Omega\cap\mathcal{I}_e$ the boundary conditions
read~$\kappa''=0$,\footnote{One can possibly ask whether the Neumann
condition~$\kappa''(\partial\mathcal{I}_p)=0$ is the optimal one or
whether it should not be replaced
with~$\kappa''(\partial\mathcal{I}_p)=c$, $0\leq c\in\mathbb{R}$. Since
this question is rather of the stability nature, it is postponed to the~\ref{Sect:B}, where stability of~$\Pi$ with respect to~$c$ will be investigated.} since~$\delta\kappa'$ has an arbitrary sign,
and~$\kappa=0$, $\kappa'=0$ as the solution~$\kappa$ with all its derivatives
vanish in~$\mathcal{I}_e$, is non-negative inside~$\mathcal{I}_p$, and is~$C^1$
continuous on~$\Omega$ by definition~\eqref{Sect:1:Eq:2c},
because~$W^{2,2}(\Omega)\subset C^1(\Omega)$, see e.g.~\cite[Section~5.6.3,
Theorem~6]{EvansPDE}.
Regarding the continuity conditions between~$\mathcal{I}_e$ and~$\mathcal{I}_p$, we infer from~\eqref{SubSect:2.1:Eq:5} that inside~$\mathcal{I}_p$, $HAl^4\kappa''$ and~$(HAl^4\kappa'')'$ must remain continuous since $\delta\kappa'$ and~$\delta\kappa$ have arbitrary signs. Outside the plastic zone, $(HAl^4\kappa'')'$ may exhibit a non-positive jump. 
}

Let us now turn our attention to the {\bf continuity or regularity conditions} that can be deduced
from the jump terms in the last two lines of  \eqref{SubSect:2.1:Eq:5}. 
Again, in the {\bf plastic domain} the variation of plastic strain and its derivative have arbitrary signs
and remain continuous, and so the jumps in $HAl^4\kappa''$ and in $(HAl^4\kappa'')'$ must vanish.
In other words, continuity of these terms must be preserved. Note that $\kappa$ and $\kappa'$ are
continuous by assumption, but continuity of $\kappa''$ or $\kappa'''$ is not assumed apriori,
and in fact is not maintained at points where for instance $H$ or $A$ have a jump. 

Inside the {\bf elastic domain}, the plastic strain is identically zero and thus all its derivatives
are zero, too, which means that the corresponding
jump terms in  the last two lines of  \eqref{SubSect:2.1:Eq:5} automatically vanish.  
However, special attention should be paid to those points of the boundary of the elastic domain
which at the same time belong to the closure of the plastic domain 
(recall that the plastic domain is an open set), i.e., to the points of
the {\bf elastoplastic
interface}, formally defined as 
$\partial\mathcal{I}_{ep} \equiv \mathcal{I}_e \cap \overline{\mathcal{I}}_p$.  
Since these are not internal points of $\mathcal{I}_e$, we cannot directly infer that all derivatives
of $\kappa$ vanish here. Continuous differentiability of $\kappa$ implies that $\kappa=0$
and $\kappa'=0$ at $\partial\mathcal{I}_{ep}$, but higher derivatives 
could in principle exhibit a jump.
So it is necessary to examine again the corresponding jump terms in \eqref{SubSect:2.1:Eq:5}.
The variation $\delta\kappa$ at $\partial\mathcal{I}_{ep}$
can be zero or positive, but never negative. Therefore, the resulting optimality condition
is the inequality $\llbracket(HAl^4\kappa'')'\rrbracket\leq 0$. 

The last condition to be derived
\new{is the most delicate one}. The derivative of the variation of plastic strain, $\delta\kappa'$,
cannot be set to a nonzero value at a point of 
$\partial\mathcal{I}_{ep}$ without simultaneously prescribing a positive
value of $\delta\kappa$, otherwise the admissibility condition $\kappa(x)+\delta\kappa(x)\ge 0$ 
would be violated in the vicinity of that point. Still, various combinations of $\delta\kappa$
and $\delta\kappa'$ can be selected such that the latter becomes increasingly ``more important''
and the jump term with $\delta\kappa$, which could potentially compensate for the negative
contribution of the jump term with $\delta\kappa'$, becomes negligible. This argument leads to
the conclusion that the jump term with $\delta\kappa'$ must vanish, i.e., $HAl^4\kappa''$
must remain continuous. Since $\kappa=0$ in $\mathcal{I}_e$ and $HAl^4\neq 0$, we must have $\kappa''=0$ 
at $\partial\mathcal{I}_{ep}$. To avoid any doubt that this optimality condition is necessary,
it is demonstrated in~\ref{Sect:B} that if the potential solutions of the localization problem are
constructed with the condition $\kappa''=0$ at $\partial\mathcal{I}_{ep}$
relaxed and then the minimum principle is imposed, 
the resulting optimum solution is the same as that constructed directly, with condition 
$\kappa''=0$  at $\partial\mathcal{I}_{ep}$ explicitly imposed.

For clarity and completeness, we compare the governing equations, boundary conditions, and regularity conditions for internal variable~$\kappa$ and the two different formulations in Tab.~\ref{SubSect:2.2:Tab:1}. Conditions for the standard solution can be found in~\cite{SoftJirZemVond} and~\cite{JirasRolshoPart2}.

To simplify the following discussions, the effect of body forces will be neglected, i.e.~$b=0$ in~\eqref{SubSect:2.1:Eq:5}, which implies that the axial force
\begin{equation}
F=EA(u'-\kappa)
\label{SubSect:2.2:Eq:5}
\end{equation}
is constant along the bar.
\begin{table}
\caption{Comparison of standard and variational formulation: governing equations, boundary conditions, and regularity conditions for internal variable~$\kappa$.}
\centering
\renewcommand{\arraystretch}{1.5}
\begin{tabular}{c|c|l}
Standard formulation & Variational formulation & \multicolumn{1}{c}{Note} \\\hline
$HA\kappa-HAl^4\kappa^\mathrm{IV}+$ & $HA\kappa-(HAl^4\kappa'')''+$ & \multirow{2}{*}{in~$\mathcal{I}_p$} \\
$+A\sigma_0=EA(u'-\kappa)$, $\kappa> 0$ & $+A\sigma_0=EA(u'-\kappa)$, $\kappa> 0$ & \\
$A\sigma_0\geq EAu'$, $\kappa=0$ & $A\sigma_0\geq EAu'$, $\kappa=0$ & in~$\mathcal{I}_e$\\
$\kappa>0$, $\kappa''=0$, $\kappa'''=0$ & $\kappa>0$, $\kappa''=0$, $(HAl^4\kappa'')'=0$ & at~$\partial\Omega\cap\mathcal{I}_p$ \\
$\kappa=0$, $\kappa''=0$ & $\kappa=0$, $\kappa''=0$, $(HAl^4\kappa'')' n\ge0$ &
at~$\partial\Omega\cap\mathcal{I}_e$ \\ continuous $\kappa$, $\kappa'$, $\kappa''$ & continuous $\kappa$, $\kappa'$, $HAl^4\kappa''$ & in~$\Omega$ \\
continuous $\kappa'''$ & continuous $(HAl^4\kappa'')'$ & in~$\Omega \setminus\partial\mathcal{I}_{ep}$ \\
$\times$ & $\llbracket(HAl^4\kappa'')'\rrbracket\leq 0$ & at
$\partial\mathcal{I}_{ep}$
\end{tabular}
\label{SubSect:2.2:Tab:1}
\end{table}
%
%-----------------------------------------------------------------------------
%	SECTION 3
%-----------------------------------------------------------------------------
%
\section{Bar With Piecewise Constant Yield Stress Distribution}
\label{Sect:3}
\begin{figure}[b]
\centering
\includegraphics[scale=1]{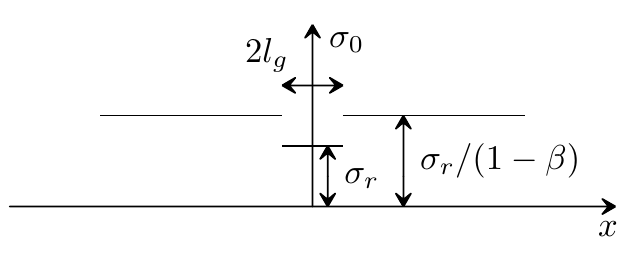}
\caption{Distribution of piecewise constant initial yield stress~$\sigma_0(x)$ of a uniform bar.}
\label{Sect:3:Fig:1}
\end{figure}

Having derived the governing equations, boundary and regularity conditions, we
now proceed to localization analysis of a tensile test of a bar with variable
initial yield stress. Let us consider a bar containing a weak segment of
length~$2l_g$ and an initial reference yield stress~$\sigma_r$, while the
remaining parts have a larger initial yield stress~$\sigma_r/(1-\beta)$ where\new{~$\beta\in(0,1)$} denotes a dimensionless parameter, cf. Fig.~\ref{Sect:3:Fig:1}. The origin of the coordinate system is placed into the centre of the weak segment, as we will consider symmetric solutions without loss of generality, i.e.~$\mathcal{I}_p=(-L_p/2,L_p/2)$ where~$L_p$ denotes the length of the plastic zone~$\mathcal{I}_p$.

Under these assumptions, the initial yield stress distribution is given by
\begin{equation}
\sigma_0(x)=\left\{\begin{array}{l l}
\sigma_r & \mbox{for }|x|<l_g \\
\sigma_r/(1-\beta) & \mbox{for }|x|>l_g
\end{array}\right.
\label{Sect:3:Eq:1}
\end{equation}
with discontinuities at~$x=\pm l_g$, cf. Fig.~\ref{Sect:3:Fig:1}. The response of the bar is elastic as long as the stress remains below the plastic limit, and the onset of yielding occurs when the yield limit is attained, i.e., when~$F=F_r$ with~$F_r=A\sigma_r=$~elastic limit force.
%
%----------------------------------
%	SUBSECTION 3.1
%----------------------------------
%
\subsection{Plastic Zone Contained in Weak Segment}
\label{SubSect:3.1}
First, let us assume that the plastic zone~$\mathcal{I}_p$ is fully contained in the weak segment, i.e.~$\mathcal{I}_p\subset (-l_g,l_g)$. Then, the yield condition in Eq.~\eqref{SubSect:2.2:Eq:2}, upon substitution of~$A(x)=A_c$ and~$EA(u'-\kappa)=F=A_c\sigma_c$, is simplified to
\begin{equation}
l^4\kappa^\mathrm{IV}(x)-\kappa(x)=\frac{\sigma_r-\sigma_c}{H}\mbox{ for }x\in\mathcal{I}_p
\label{SubSect:3.1:Eq:1}
\end{equation}
which is a fourth-order linear differential equation with constant coefficients and a constant right-hand side. It will be convenient for further analysis to convert Eq.~\eqref{SubSect:3.1:Eq:1} into a normalized form. To this purpose, we introduce
\begin{itemize}
	\item dimensionless spatial coordinate~$\xi=x/l$,
	\item plastic strain at complete failure for the local model~$\kappa_f=-\sigma_r/H$,
	\item normalized plastic strain~$\kappa_n=\kappa/\kappa_f=-H\kappa/\sigma_r$,
	\item dimensionless stress or load parameter~$\phi=F/F_r$,
	\item dimensionless parameters describing the ratio~$\lambda_g=l_g/l$ between "geometric" and material characteristic lengths, and
	\item ratio~$\lambda_p=L_p/2l$ between one half of plastic zone and the material characteristic length.
\end{itemize}
In terms of normalized quantities, Eq.~\eqref{SubSect:3.1:Eq:1} is transformed into
\begin{equation}
\kappa_n^\mathrm{IV}(\xi)-\kappa_n(\xi)=\phi-1\mbox{ for }\xi\in(-\lambda_p,\lambda_p)
\label{SubSect:3.1:Eq:2}
\end{equation}
where, for simplicity, the derivatives with respect to~$\xi$ are still denoted
by primes or Latin numerals. The general solution to
Eq.~\eqref{SubSect:3.1:Eq:2} is
\begin{equation}
\kappa_n(\xi)=C_1\cos{\xi}+C_2\sin{\xi}+C_3\cosh{\xi}+C_4\sinh{\xi}+1-\phi
\label{SubSect:3.1:Eq:3}
\end{equation}
where the integration constants~$C_2$ and~$C_4$ vanish due to the symmetry conditions~$\kappa_n'(0)=0,\kappa_n'''(0)=0$. The remaining unknowns are the integration constants~$C_1$, $C_3$, and the size of the plastic zone~$\lambda_p$, which are determined from the regularity conditions at the boundary of the plastic zone~$\partial\mathcal{I}_p$,
\begin{equation}
\kappa_n(\lambda_p)=0,\quad\kappa_n'(\lambda_p)=0,\quad\kappa_n''(\lambda_p)=0
\label{SubSect:3.1:Eq:4}
\end{equation}
Substitution of the general solution~\eqref{SubSect:3.1:Eq:3} into the boundary conditions~\eqref{SubSect:3.1:Eq:4} leads to the set of three equations
\begin{equation}
\begin{aligned}
C_1\cos{\lambda_p}+C_3\cosh{\lambda_p}&=\phi-1\\
C_1\sin{\lambda_p}&=C_3\sinh{\lambda_p}\\
C_1\cos{\lambda_p}&=C_3\cosh{\lambda_p}
\end{aligned}
\label{SubSect:3.1:Eq:5}
\end{equation}
Elimination of~$C_1$ and~$C_3$ reduces the above system to a single nonlinear equation 
\begin{equation}
\tan{\lambda_p}=\tanh{\lambda_p}
\label{SubSect:3.1:Eq:6}
\end{equation}
\new{Solutions of this transcendental equation can be found numerically;
let us denote them \mbox{$\pm m_i\doteq 0,\pm
3.9266,\pm 7.0686,\dots,i\in\mathbb{N}_0$}. Of course, only positive values
are of physical interest. For given $i\in\mathbb{N}$, the integration
constants can be evaluated from the first two equations of the system
\eqref{SubSect:3.1:Eq:5} as
\begin{equation}
\begin{aligned}
C_1&=\frac{\phi-1}{2\cos m_i}\\
C_3&=\frac{\phi-1}{2\cosh m_i}
\end{aligned}
\label{SubSect:3.1:Eq:7}
\end{equation}
The most localized plastic strain profile is obtained for $i=1$, i.e., for
$\lambda_p=m_1$.}
This is also the standard, non-variational solution of the localization problem
for a bar with perfectly uniform properties presented by~\cite{JirasRolshoPart2}, Eq.~(40). 

Outside the plastic zone, condition~\eqref{SubSect:2.2:Eq:4} simplifies to
\begin{equation}
\sigma_c\leq\sigma_r \mbox{ or, equivalently, } \phi\leq 1
\label{SubSect:3.1:Eq:8}
\end{equation}
Analysis of the second variation, presented in detail in~\ref{Sect:A}, 
reveals that the most localized solution is stable \new{(provided that the material parameters and bar length 
satisfy a certain condition)}, i.e.~it corresponds to a local minimum of~$\Pi$, \new{
while the solutions for $i\ge 2$ are unstable.}

So far we have assumed that the weak segment is long enough, $2l_g>2lm_1$, and thus
the stable localized solution is not affected by stronger parts of the bar. Nevertheless, if the weak segment is shorter, the analysis needs to be modified.
%
%----------------------------------
%	SUBSECTION 3.2
%----------------------------------
%
\subsection{Plastic Zone Extending to Strong Segments}
\label{SubSect:3.2}
\begin{figure}
\centering
\includegraphics[scale=1]{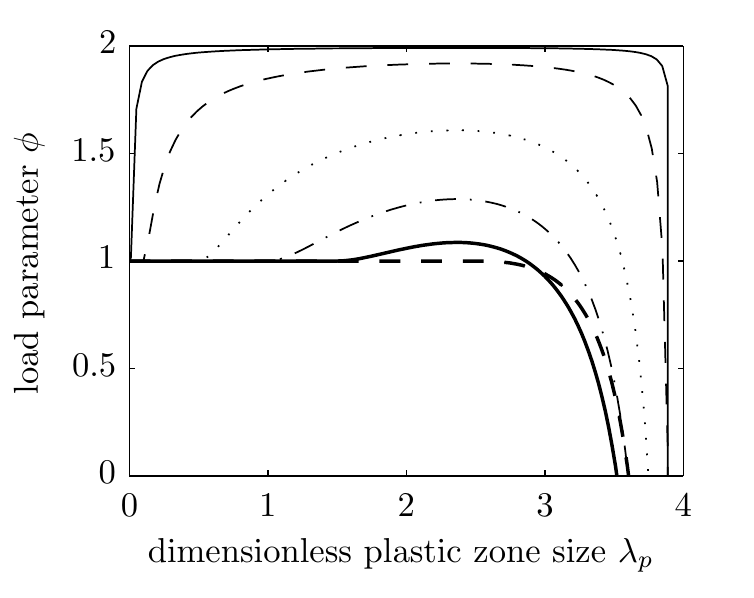}
\caption{Piecewise constant yield stress distribution: relation between load
parameter and plastic zone size for several values of~$\lambda_g$, and
for~$\beta=0.5$; for complete legend please refer to
Fig.~$\ref{SubSect:3.2:Fig:3a}$.}
\label{SubSect:3.2:Fig:1}
\end{figure}

Let us proceed to the case when~$L_p>2l_g$, i.e.~$\lambda_g<m_1$. In this situation, Eq.~\eqref{SubSect:3.1:Eq:2} must be extended to the parts surrounding the weak segment:
\begin{equation}
\begin{aligned}
\kappa_n^\mathrm{IV}(\xi)-\kappa_n(\xi)&=\phi-1&\mbox{ for }&&-\lambda_g<\xi<\lambda_g&\\
\kappa_n^\mathrm{IV}(\xi)-\kappa_n(\xi)&=\phi+\frac{1}{\beta-1}&\mbox{ for }&&\lambda_g<|\xi|<\lambda_p&\\
\end{aligned}
\label{SubSect:3.2:Eq:1}
\end{equation}
yielding the general solution
\begin{equation}
\kappa_n(\xi)=\left\{
\begin{aligned}
& C_1\cos{\xi}+C_2\sin{\xi}+C_3\cosh{\xi}+C_4\sinh{\xi}+1-\phi & \mbox{for }&& -\lambda_g<\xi<\lambda_g&\\
& C_5\cos{\xi}+C_6\sin{\xi}+C_7\cosh{\xi}+C_8\sinh{\xi}+\frac{1}{1-\beta}-\phi & \mbox{for } && \lambda_g<\xi<\lambda_p&\\
& C_9\cos{\xi}+C_{10}\sin{\xi}+C_{11}\cosh{\xi}+C_{12}\sinh{\xi}+\frac{1}{1-\beta}-\phi & \mbox{for } && -\lambda_p<\xi<\lambda_g&
\end{aligned}\right.
\label{SubSect:3.2:Eq:2}
\end{equation}
By the symmetry conditions, the integration constants~$C_2$ and~$C_4$ vanish again, and
\begin{equation}
\begin{aligned}
C_5&=C_{9},\\
C_6&=-C_{10},\\
C_7&=C_{11},\\
C_8&=-C_{12}
\end{aligned}
\label{SubSect:3.2:Eq:2a}
\end{equation}
The remaining unknown constants~$C_i$ for~$i=1,3,5,6,7,8$ and the
dimensionless plastic zone size~$\lambda_p$ can be determined from seven
conditions; namely from continuity of~$\kappa$, $\kappa'$, $HAl^4\kappa''$,
and~$(HAl^4\kappa'')'$ at~$\xi=\lambda_g$, and of~$\kappa$, $\kappa'$,
and~$HAl^4\kappa''$ at~$\xi=\lambda_p$. Because the resulting set of equations
is nonlinear in~$\lambda_p$, it is more convenient to solve the system for~$C_i$
and~$\phi$, with~$\lambda_p$ considered as given. 
In other words, the loading process is considered as parametrized by $\lambda_p$ instead of $\phi$. 
Then, we arrive at a set of
seven linear equations in the form
\begin{equation}
\begin{aligned}&
\left(\begin{array}{c c r r}
-\cos{\lambda_g} & -\cosh{\lambda_g} & \cos{\lambda_g} & \sin{\lambda_g} \\
\sin{\lambda_g} & -\sinh{\lambda_g} & -\sin{\lambda_g} & \cos{\lambda_g} \\
-\cos{\lambda_g} & \cosh{\lambda_g} & \cos{\lambda_g} & \sin{\lambda_g} \\
\sin{\lambda_g} & \sinh{\lambda_g} & -\sin{\lambda_g} & \cos{\lambda_g} \\
0 & 0 & \cos{\lambda_p} & \sin{\lambda_p} \\
0 & 0 & -\sin{\lambda_p} & \cos{\lambda_p} \\
0 & 0 & \cos{\lambda_p} & \sin{\lambda_p} 
\end{array}\right.
\\&\qquad\qquad\qquad
\left.\begin{array}{r r c}
\cosh{\lambda_g} & \sinh{\lambda_g} & 0 \\
\sinh{\lambda_g} & \cosh{\lambda_g} & 0 \\
-\cosh{\lambda_g} & -\sinh{\lambda_g} & 0 \\
-\sinh{\lambda_g} & -\cosh{\lambda_g} & 0 \\
\cosh{\lambda_p} & \sinh{\lambda_p} & -1 \\
\sinh{\lambda_p} & \cosh{\lambda_p} & 0 \\
-\cosh{\lambda_p} & -\sinh{\lambda_p} & 0 
\end{array}\right)
\left(\begin{array}{c}
C_1 \\ C_3 \\ C_5 \\ C_6 \\ C_7 \\ C_8 \\ \phi
\end{array}\right)=
\left(\begin{array}{c}
\frac{\beta}{\beta-1} \\ 0 \\ 0 \\ 0 \\ \frac{1}{\beta-1} \\ 0 \\ 0
\end{array}\right)
\end{aligned}
\label{SubSect:3.2:Eq:4}
\end{equation}
which can be easily solved by matrix inversion; for the sake of brevity, however, we do not provide the results in the explicit form. The resulting dependencies between the load parameter~$\phi$ and normalized plastic zone~$\lambda_p$ are depicted in Fig.~\ref{SubSect:3.2:Fig:1} for several values of~$\lambda_g=m_1\{0.003,0.025,0.127,0.255,0.382,0.637\}$ and for~$\beta=0.5$.

\begin{figure}
\centering
\subfloat[]{\includegraphics[scale=1]{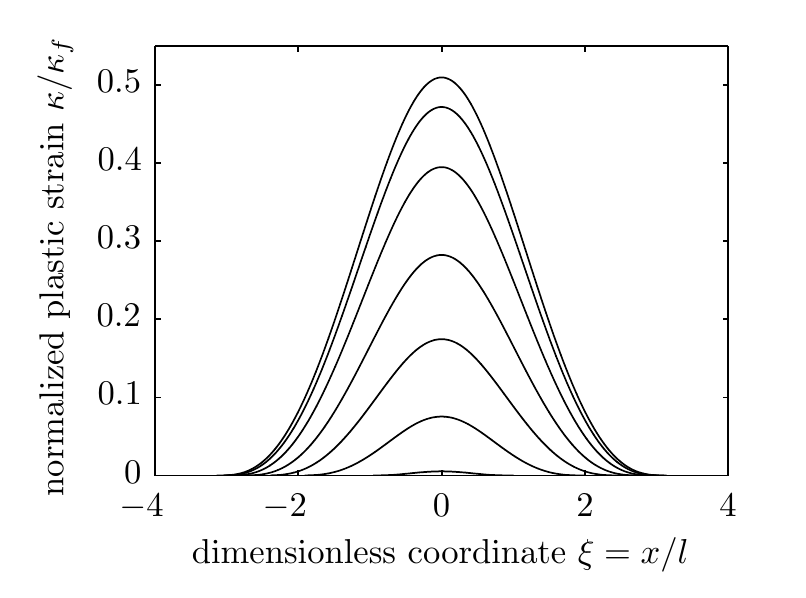}\label{SubSect:3.2:Fig:2a}}
\subfloat[]{\includegraphics[scale=1]{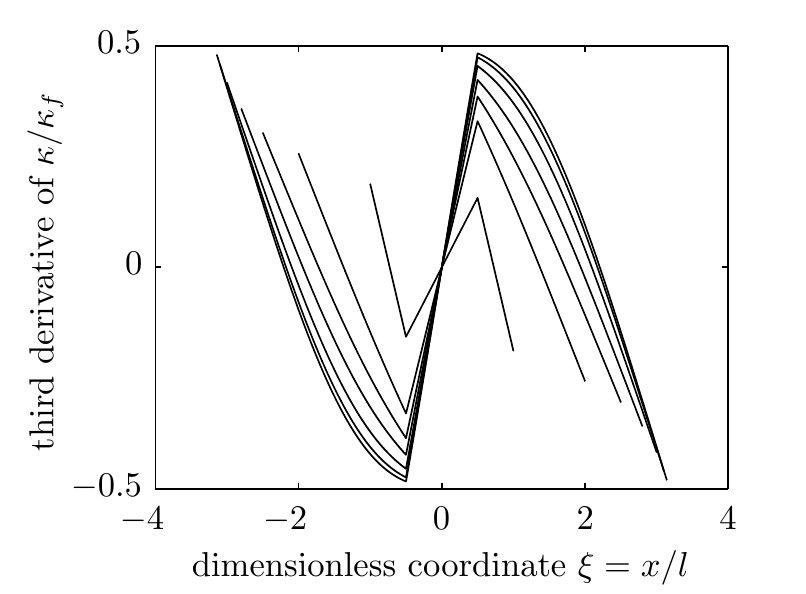}\label{SubSect:3.2:Fig:2b}}
\caption{Piecewise constant yield stress distribution: (a) evolution of plastic strain profile and (b) its third derivative for monotonically increasing plastic zone length~$\lambda_p$.}
\label{SubSect:3.2:Fig:2}
\end{figure}

Solving the system in Eq.~\eqref{SubSect:3.2:Eq:4} and \new{substituting} the results into Eq.~\eqref{SubSect:3.2:Eq:2} leads to the distribution of plastic strain~$\kappa_n$. An example for~$\lambda_g=0.127m_1$, $\beta=0.5$ and the monotonically expanding plastic zone~$\lambda_p=m_1\{0.277,0.554,0.693,0.776,0.831,0.858,0.870\}$ is shown in Fig.~\ref{SubSect:3.2:Fig:2a}. In Fig.~\ref{SubSect:3.2:Fig:2b}, the third derivative of plastic strain is depicted, satisfying all the regularity requirements summarized in Tab.~\ref{SubSect:2.2:Tab:1}.

\begin{figure}
\centering
\subfloat[]{\includegraphics[scale=1]{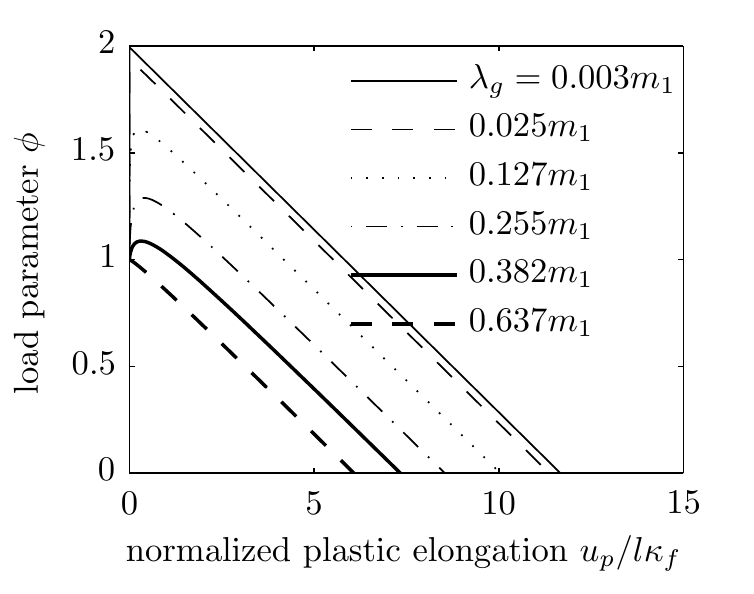}\label{SubSect:3.2:Fig:3a}}
\subfloat[]{\includegraphics[scale=1]{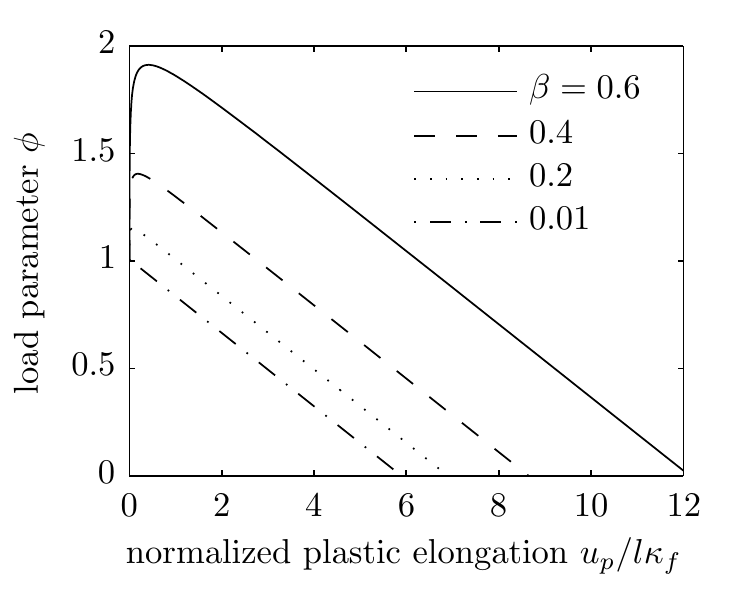}\label{SubSect:3.2:Fig:3b}}
\caption{Piecewise constant yield stress distribution: (a) plastic part of dimensionless load-displacement diagram for different values of dimensionless imperfection length~$\lambda_g$ assuming fixed~$\beta=0.5$, and~(b) for different values of dimensionless imperfection magnitude~$\beta$ assuming fixed~$\lambda_g=0.127m_1$.}
\label{SubSect:3.2:Fig:3}
\end{figure}

Integrating Eq.~\eqref{SubSect:3.2:Eq:2} over the length of the plastic zone provides the normalized plastic elongation
\begin{equation}
\frac{u_p}{l\kappa_f}=\int_{-\lambda_p}^{\lambda_p}\kappa_n(\xi)\,\mathrm{d}\xi=2\int_{0}^{\lambda_g}\kappa_n(\xi)\,\mathrm{d}\xi+2\int_{\lambda_g}^{\lambda_p}\kappa_n(\xi)\,\mathrm{d}\xi
\label{SubSect:3.2:Eq:5}
\end{equation}
which in this simple case can be carried out analytically:
\begin{equation}
\begin{aligned}
\frac{u_p}{l\kappa_f}=&\, 2\bigg\{\frac{\beta\lambda_g+\lambda_p[\phi(1-\beta)-1]}{\beta-1}+C_1\sin{\lambda_g}+C_3\sinh{\lambda_g}+C_5(\sin{\lambda_p}-\sin{\lambda_g})\\
&+C_6(\cos{\lambda_g}-\cos{\lambda_p})+C_7(\sinh{\lambda_p}-\sinh{\lambda_g})+C_8(\cosh{\lambda_p}-\cosh{\lambda_g})\bigg\}
\end{aligned}
\label{SubSect:3.2:Eq:6}
\end{equation}
The dimensionless load-plastic elongation diagrams for fixed~$\beta=0.5$ and different dimensionless sizes of the weak segment~$\lambda_g$ are shown in Fig.~\ref{SubSect:3.2:Fig:3a}, and for fixed~$\lambda_g=0.127m_1$ with different values of~$\beta$ in Fig.~\ref{SubSect:3.2:Fig:3b}. These figures reflect the influence of both parameters on the shape of the load-displacement curve for a bar with an imperfection, where~$\lambda_g$ is understood as the length of the imperfection, and~$\beta$ as its magnitude. Note that regardless of the imperfection shape and size, the plastic response of the bar is always of the same slope. For shorter imperfections we observe significant hardening, while for longer imperfections the response attains the behaviour of a perfectly uniform bar. The imperfection magnitude influences the load-displacement diagram in the opposite way; for large values of~$\beta$, the response exhibits a higher maximum force. On the other hand, in the case of small magnitudes the response approaches the behaviour of a uniform bar discussed in Section~\ref{SubSect:3.1}. As an alternative physical interpretation, we could consider a semi-infinite layer of material between two parallel planes which are mutually displaced in tangential direction and left unconstrained in the normal direction, so that the layer with vertically variable material properties is under pure shear stress.

In order to demonstrate that we have obtained an admissible solution, we check condition~\eqref{SubSect:2.2:Eq:4}, which is valid outside the plastic zone and simplifies to
\begin{equation}
\phi\leq\frac{1}{1-\beta}
\label{SubSect:3.2:Eq:7}
\end{equation}
The condition can be simply verified in Fig.~\ref{SubSect:3.2:Fig:1}, where we have~$\phi\leq 2$ for~$\beta=0.5$.

\begin{figure}
\centering
\includegraphics[scale=1]{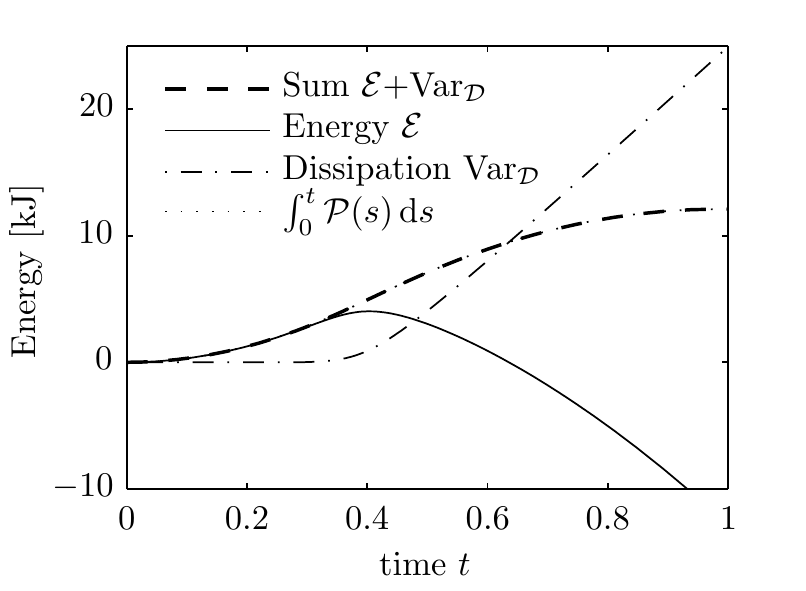}
\caption{Piecewise constant yield stress distribution: energy evolution for~$\beta=0.5$; see Tab.~\ref{SubSect:3.2:Tab:1} for the physical parameters and
Eq.~\eqref{SubSect:3.2:Eq:8} for the loading program.}
\label{SubSect:3.2:Fig:4}
\end{figure}

Finally, in Fig.~\ref{SubSect:3.2:Fig:4}, we check that the energy
balance~\eqref{Sect:1:Eq:E} holds along the whole loading process in
agrement with general results~\cite[Section~2.2.3]{Pham:2011:IUS}
and~\cite[Property~1]{Pham:2013:ODM} for solutions sufficiently regular in time. We
observe that the response is first elastic, i.e.~the~$\mathcal{E}$ curve is quadratic with no dissipation~$\mbox{Var}_\mathcal{D}$,
followed by the evolution of the plastic strain accompanied by nonzero
dissipation. For this graph, physical constants presented in
Tab.~\ref{SubSect:3.2:Tab:1} were used. Parameters that control the imperfection were set to~$\lambda_g=0.255m_1$ and~$\beta=0.5$, the total length
of the bar was~$L=4lm_1$, and the evolution was parametrized by the
dimensionless plastic zone length
\begin{equation}
\lambda_p(t)=\lambda_g+t(m_1-\lambda_g)\mbox{ for }t\in[0,1]
\label{SubSect:3.2:Eq:8}
\end{equation}
\begin{table}
\caption{Physical and geometric parameters of all test examples.}
\centering
\renewcommand{\arraystretch}{1.5}
\begin{tabular}{l|r@{}l}
\multicolumn{1}{c|}{Physical parameters} & \multicolumn{2}{c}{Values} \\\hline
Young's modulus, $E$ & $60$ & ~GPa \\
Softening modulus, $H$ & $-10$ & ~GPa \\
Characteristic length, $l$ & $0$ & $.05$~m \\
Reference initial yield stress, $\sigma_r$ & $200$ & ~MPa \\
Weakest cross-sectional area, $A_c$ & $0$ & $.01$~m$^2$ \\
\end{tabular}
\label{SubSect:3.2:Tab:1}
\end{table}
%
%-----------------------------------------------------------------------------
%	SECTION 4
%-----------------------------------------------------------------------------
%
\section{Bar With Piecewise Constant Stress Distribution}
\label{Sect:4}
\begin{figure}
\centering
\subfloat[]{\includegraphics[scale=1]{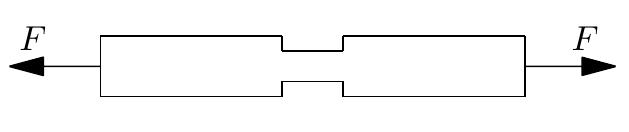}\label{Sect:4:Fig:1a}}
\subfloat[]{\includegraphics[scale=1]{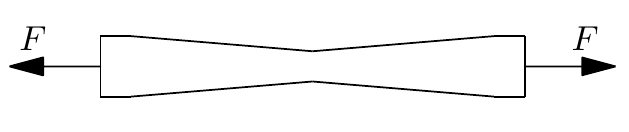}\label{Sect:4:Fig:1b}}\\
\subfloat[]{\includegraphics[scale=1]{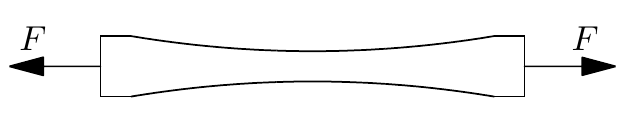}\label{Sect:4:Fig:1c}}
\caption{Geometries of tensile test bars corresponding to (a)~discontinuous, (b)~continuous, but not continuously differentiable, and (c)~infinitely smooth stress fields.}
\label{Sect:4:Fig:1}
\end{figure}

In all subsequent sections, contrary to the previous one, we will assume the initial yield stress to be constant,  i.e.~$\sigma_0(x)=\sigma_r$, and will investigate the influence of the variable cross-sectional area resulting in spatially variable stress field distributions. As the first example, let us consider a very similar load test to that presented in Section~\ref{Sect:3}, but now with discontinuous sectional area, i.e.~a bar containing a thin segment of length~$2l_g$ and sectional area~$A_c$, and with remaining thick parts of sectional area~$A_c/(1-\beta)$ where, as previously, \new{$\beta\in(0,1)$} denotes a dimensionless parameter, cf. Fig.~\ref{Sect:4:Fig:1a}.

\begin{figure}
\centering
\includegraphics[scale=1]{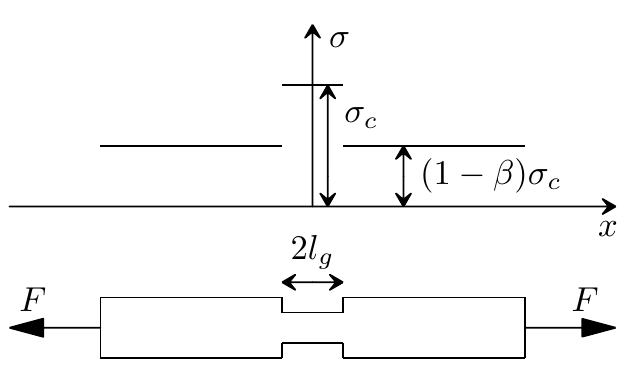}
\caption{Bar with piecewise constant cross-sectional area and the corresponding stress distribution.}
\label{Sect:4:Fig:2}
\end{figure}

The corresponding stress distribution is described by
\begin{equation}
\sigma(x)=\left\{\begin{array}{l l}
F/A_c=\sigma_c & \mbox{for }|x|<l_g \\
F/[A_c/(1-\beta)]=(1-\beta)\sigma_c & \mbox{for }|x|>l_g
\end{array}\right.
\label{Sect:4:Eq:1}
\end{equation}
again with discontinuities at~$x=\pm l_g$, cf. Fig.~\ref{Sect:4:Fig:2}. Clearly, the case of a plastic zone contained in the \new{weak} segment coincides exactly with the situation presented in Section~\ref{SubSect:3.1}, and hence is not discussed again. Instead, let us proceed directly to the case in which~$L_p>2l_g$, i.e., $\lambda_g\leq m_1$.
%
%----------------------------------
%	SUBSECTION 4.1
%----------------------------------
%
\subsection{Plastic Zone Extending to Thick Segments}
\label{SubSect:4.1}
Substituting Eq.~\eqref{Sect:4:Eq:1} into the yield
condition~\eqref{SubSect:2.2:Eq:2} and converting the result into the
dimensionless form, we obtain the governing equations, cf. Eq.~\eqref{SubSect:3.2:Eq:1}:
\begin{equation}
\begin{aligned}
\kappa_n^\mathrm{IV}(\xi)-\kappa_n(\xi)&=\phi-1&\mbox{ for }&&-\lambda_g<\xi<\lambda_g&\\
\kappa_n^\mathrm{IV}(\xi)-\kappa_n(\xi)&=\phi-\beta\phi-1&\mbox{ for }&&\lambda_g<|\xi|<\lambda_p&\\
\end{aligned}
\label{SubSect:4.1:Eq:1}
\end{equation}
which provide the general solution
\begin{equation}
\kappa_n(\xi)=\left\{
\begin{aligned}
& C_1\cos{\xi}+C_2\sin{\xi}+C_3\cosh{\xi}+C_4\sinh{\xi}+1-\phi & \mbox{for }&& -\lambda_g<\xi<\lambda_g&\\
& C_5\cos{\xi}+C_6\sin{\xi}+C_7\cosh{\xi}+C_8\sinh{\xi}+1-\phi+\beta\phi & \mbox{for } && \lambda_g<\xi<\lambda_p&\\
& C_9\cos{\xi}+C_{10}\sin{\xi}+C_{11}\cosh{\xi}+C_{12}\sinh{\xi}+1-\phi+\beta\phi & \mbox{for } && -\lambda_p<\xi<\lambda_g&
\end{aligned}\right.
\label{SubSect:4.1:Eq:2}
\end{equation}
By the symmetry arguments, the integration constants~$C_2$ and~$C_4$ are zero, and for~$C_i$, $i=9,\dots,12$, the relations~\eqref{SubSect:3.2:Eq:2a} hold again. The remaining unknown constants~$C_i$ for~$i=1,3,5,6,7,8$ and the
dimensionless plastic zone size~$\lambda_p$ can be determined from the regularity conditions; for example, continuity
of~$HAl^4\kappa''$ and~$(HAl^4\kappa'')'$ at~$\xi=\lambda_g$ give
\begin{equation}
\begin{aligned}
(1-\beta)\kappa_n''(\lambda_g^-)&=\kappa_n''(\lambda_g^+)\\ (1-\beta)\kappa_n'''(\lambda_g^-)&=\kappa_n'''(\lambda_g^+)
\end{aligned}
\label{SubSect:4.1:Eq:3}
\end{equation}

\begin{figure}
\centering
\includegraphics[scale=1]{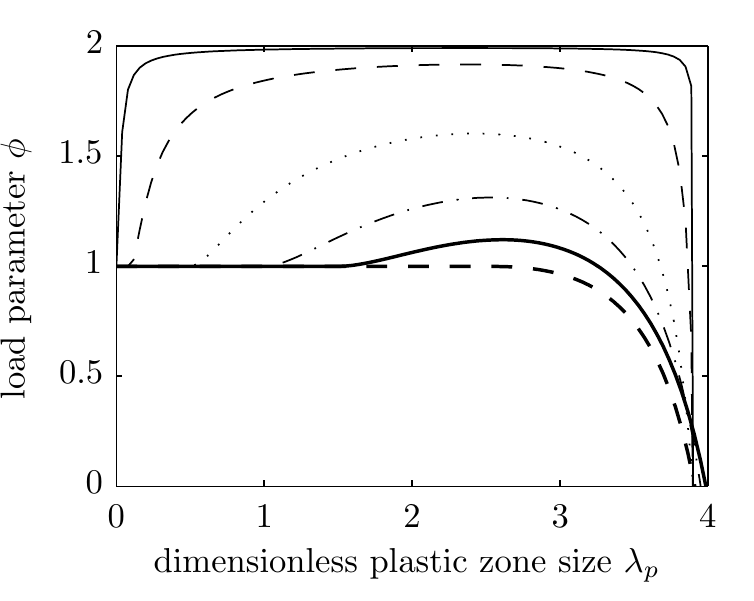}
\caption{Piecewise constant stress distribution: relation between load parameter and plastic zone size for several values of~$\lambda_g$, and for~$\beta=0.5$; for complete legend please refer to Fig.~$\ref{SubSect:4.1:Fig:3a}$.}
\label{SubSect:4.1:Fig:1}
\end{figure}

The set of seven linear equations arising from the regularity and boundary conditions reads
\begin{equation}
\begin{aligned}&
\left(\begin{array}{c c r r}
-\cos{\lambda_g} & -\cosh{\lambda_g} & \cos{\lambda_g} & \sin{\lambda_g} \\
\sin{\lambda_g} & -\sinh{\lambda_g} & -\sin{\lambda_g} & \cos{\lambda_g} \\
(\beta-1)\cos{\lambda_g} & (1-\beta)\cosh{\lambda_g} & \cos{\lambda_g} & \sin{\lambda_g} \\
(1-\beta)\sin{\lambda_g} & (1-\beta)\sinh{\lambda_g} & -\sin{\lambda_g} & \cos{\lambda_g} \\
0 & 0 & \cos{\lambda_p} & \sin{\lambda_p} \\
0 & 0 & -\sin{\lambda_p} & \cos{\lambda_p} \\
0 & 0 & \cos{\lambda_p} & \sin{\lambda_p} 
\end{array}\right.
\\&\qquad\qquad\qquad
\left.\begin{array}{r r c}
\cosh{\lambda_g} & \sinh{\lambda_g} & \beta \\
\sinh{\lambda_g} & \cosh{\lambda_g} & 0 \\
-\cosh{\lambda_g} & -\sinh{\lambda_g} & 0 \\
-\sinh{\lambda_g} & -\cosh{\lambda_g} & 0 \\
\cosh{\lambda_p} & \sinh{\lambda_p} & \beta-1 \\
\sinh{\lambda_p} & \cosh{\lambda_p} & 0 \\
-\cosh{\lambda_p} & -\sinh{\lambda_p} & 0 
\end{array}\right)
\left(\begin{array}{c}
C_1 \\ C_3 \\ C_5 \\ C_6 \\ C_7 \\ C_8 \\ \phi
\end{array}\right)=
\left(\begin{array}{c}
0 \\ 0 \\ 0 \\ 0 \\ -1 \\ 0 \\ 0
\end{array}\right)
\end{aligned}
\label{SubSect:4.1:Eq:4}
\end{equation}
and for convenience it is again solved numerically with no explicit expressions
presented. The resulting dependencies between the load parameter~$\phi$ and the normalized plastic zone size~$\lambda_p$ are depicted in Fig.~\ref{SubSect:4.1:Fig:1} for the same values~$\lambda_g$ and~$\beta$ as in Section~\ref{SubSect:3.2}, see also Fig.~\ref{SubSect:3.2:Fig:1}.

\begin{figure}
\centering
\subfloat[]{\includegraphics[scale=1]{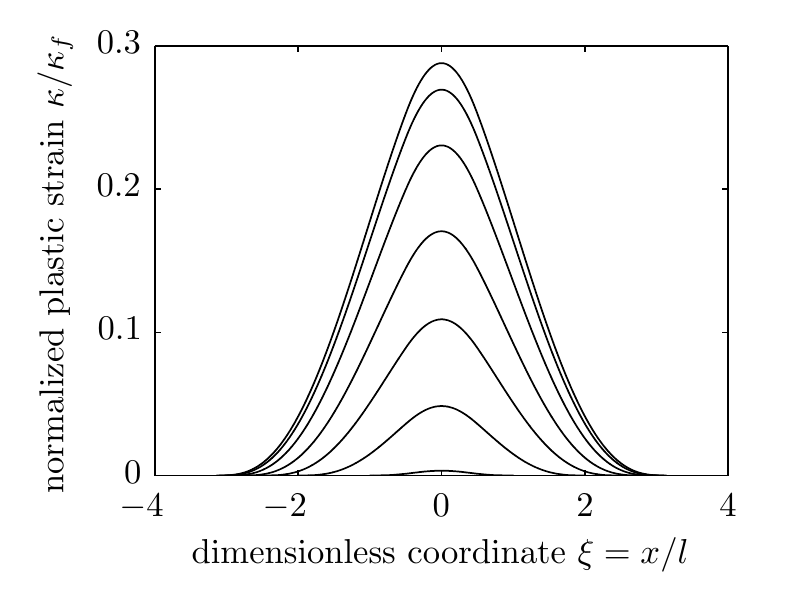}\label{SubSect:4.1:Fig:2a}}
\subfloat[]{\includegraphics[scale=1]{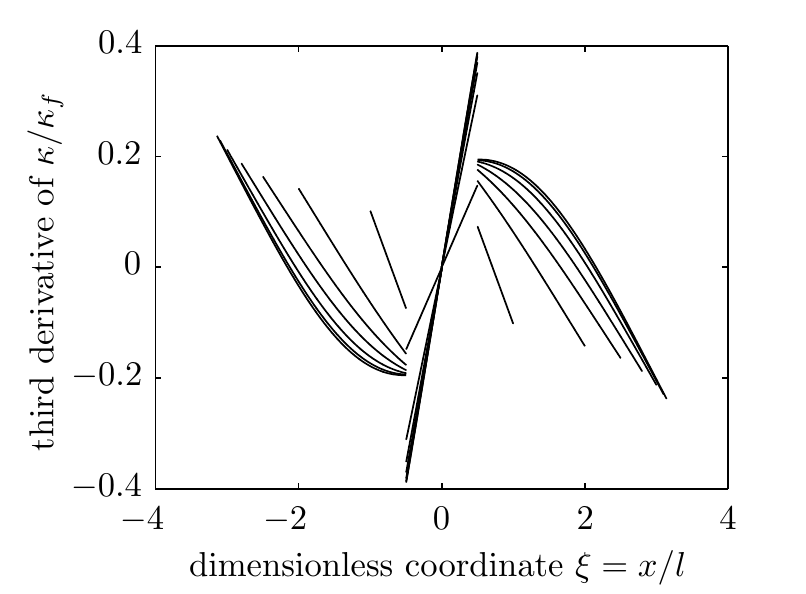}\label{SubSect:4.1:Fig:2b}}
\caption{Piecewise constant stress distribution: (a) evolution of plastic strain profile and (b) its third derivative for monotonically increasing plastic zone length~$\lambda_p$.}
\label{SubSect:4.1:Fig:2}
\end{figure}

\begin{figure}
	\centering
	\subfloat[]{\includegraphics[scale=1]{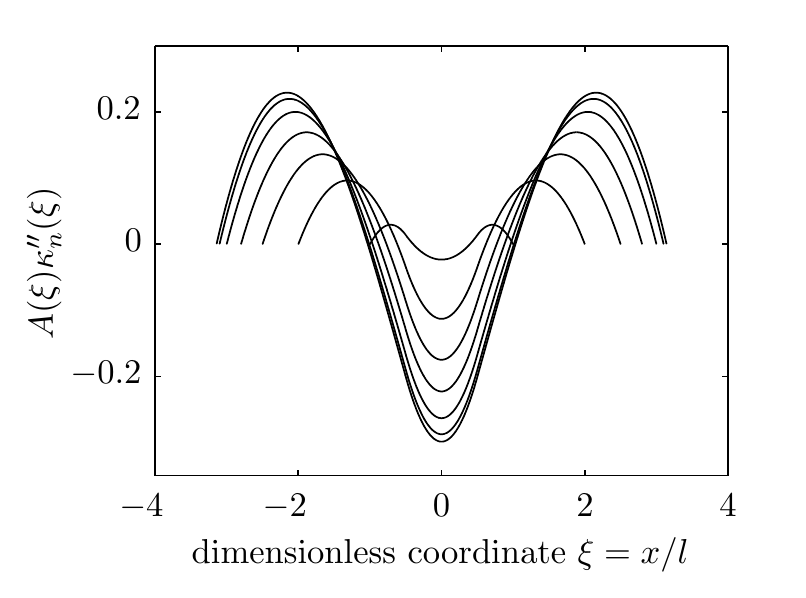}\label{SubSect:4.1:Fig:2.1a}}
	\subfloat[]{\includegraphics[scale=1]{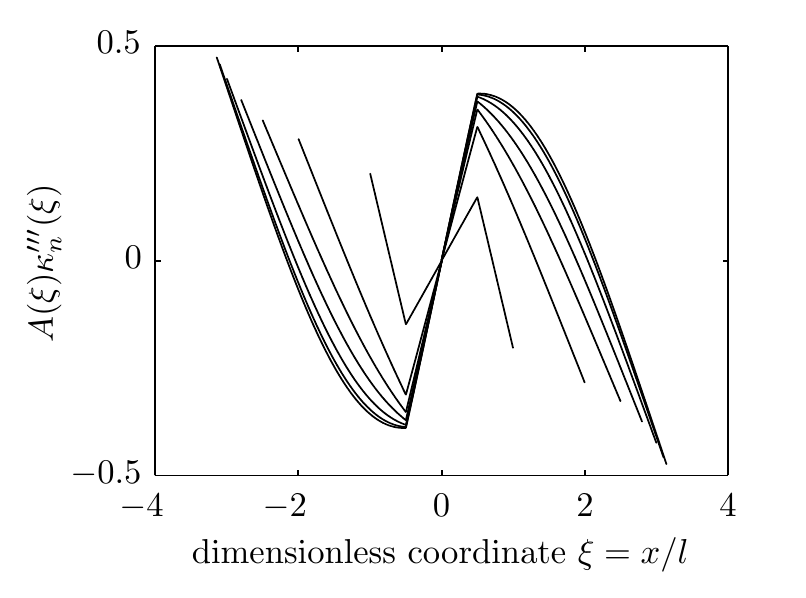}\label{SubSect:4.1:Fig:2.1b}}
	\caption{Piecewise constant stress distribution: (a) evolution of~$A(\xi)\kappa_n''(\xi)$ and (b) $A(\xi)\kappa_n'''(\xi)$ for monotonically increasing plastic zone length~$\lambda_p$, and for fixed~$\lambda_g = 0.127m_1$, $\beta=0.5$.}
	\label{SubSect:4.1:Fig:2.1}
\end{figure}

\begin{figure}
	\centering
	\includegraphics[scale=1]{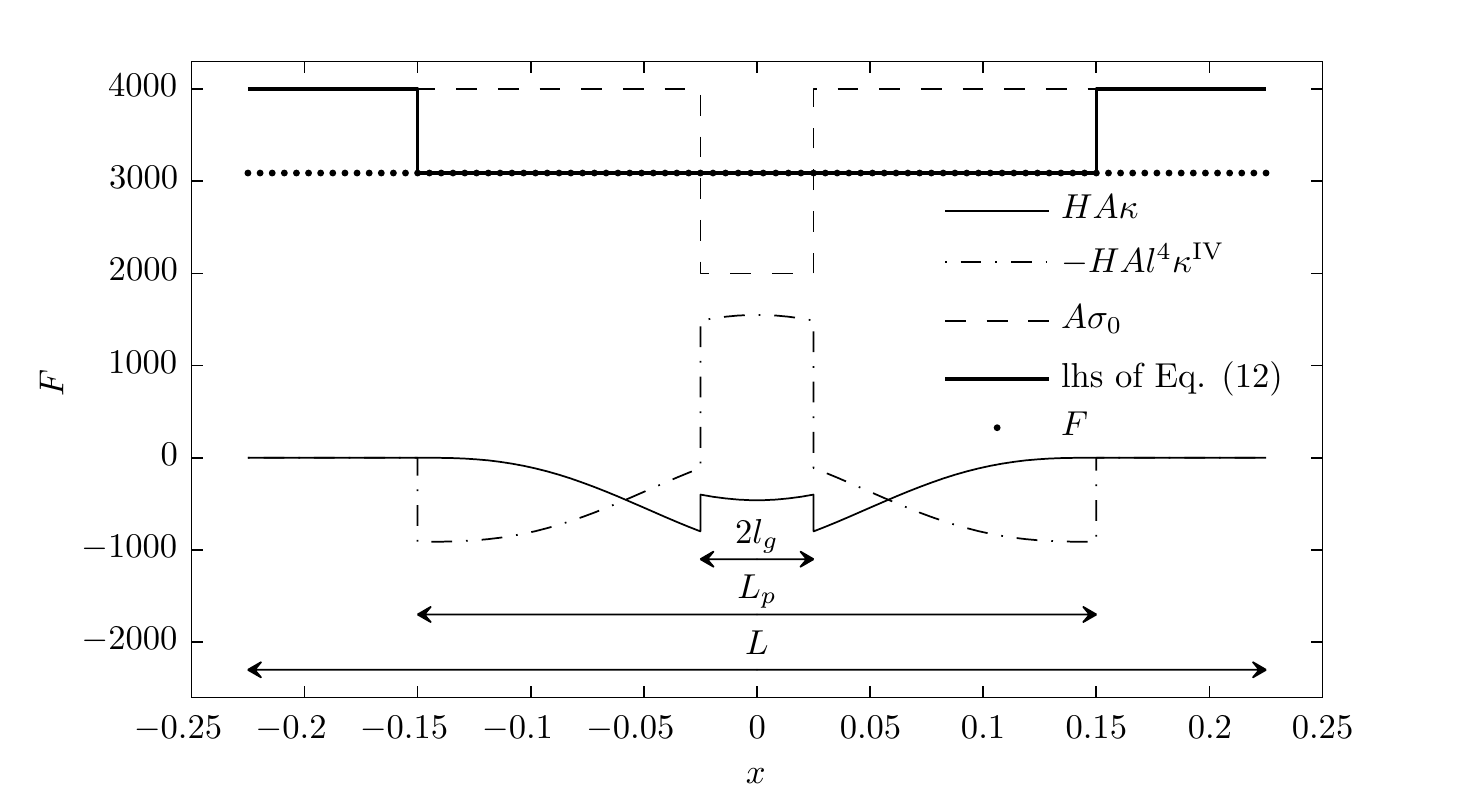}
	\caption{Piecewise constant stress distribution: a pictorial demonstration of plastic yield condition~\eqref{SubSect:2.2:Eq:2} valid in~$\mathcal{I}_p$, and plastic admissibility condition~\eqref{SubSect:2.2:Eq:3} valid in~$\mathcal{I}_e$ for~$\lambda_g=0.127m_1$, $\lambda_p=0.762m_1$, $\beta=0.5$, and~$\frac{L}{2l}=1.5\lambda_p$.}
	\label{SubSect:4.1:Fig:2.2}
\end{figure}

The normalized plastic strain, analytically expressed in Eq.~\eqref{SubSect:4.1:Eq:2}, is depicted in Fig.~\ref{SubSect:4.1:Fig:2a} for~$\lambda_g=0.127m_1$, $\beta=0.5$, and for a monotonically expanding plastic zone~$\lambda_p=m_1\{0.277,0.554,0.693,0.776,0.831,0.858,0.870\}$. Its third derivative, presented in Fig.~\ref{SubSect:4.1:Fig:2b}, exhibits discontinuities at~$\xi=\pm\lambda_g$ and~$\xi=\pm\lambda_p$. Let us note, however, that the quantity~$A(\xi)\kappa_n'''(\xi)$ plotted in Fig.~\ref{SubSect:4.1:Fig:2.1b} has non-negative jumps only at~$\xi=\pm\lambda_p$ and remains continuous for~$\xi\in(-\lambda_p,\lambda_p)$ in accordance to the discussion presented at the end of Section~\ref{SubSect:2.1}. For completeness, continuity of~$A(\xi)\kappa_n''(\xi)$ and validity of plastic yield condition~\eqref{SubSect:2.2:Eq:2} or plastic admissibility condition~\eqref{SubSect:2.2:Eq:3} can be verified in Figs.~\ref{SubSect:4.1:Fig:2.1a} and~\ref{SubSect:4.1:Fig:2.2}. 

\begin{figure}
\centering
\subfloat[]{\includegraphics[scale=1]{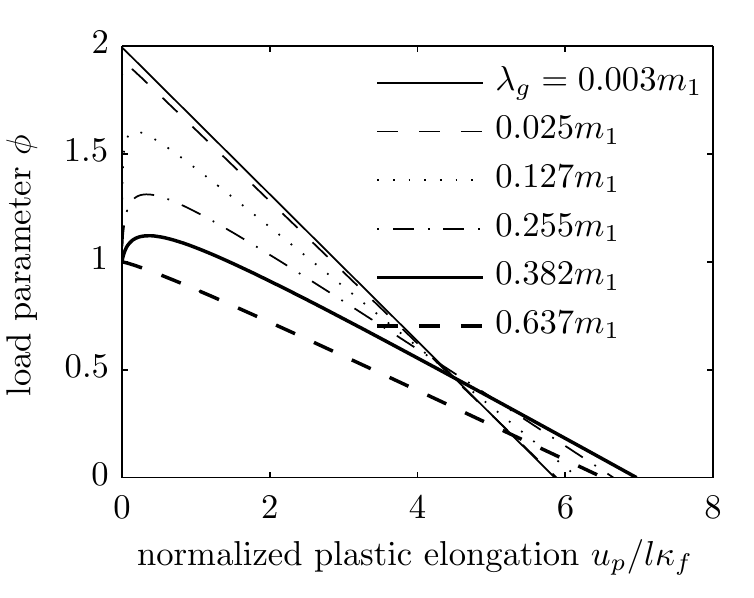}\label{SubSect:4.1:Fig:3a}}
\subfloat[]{\includegraphics[scale=1]{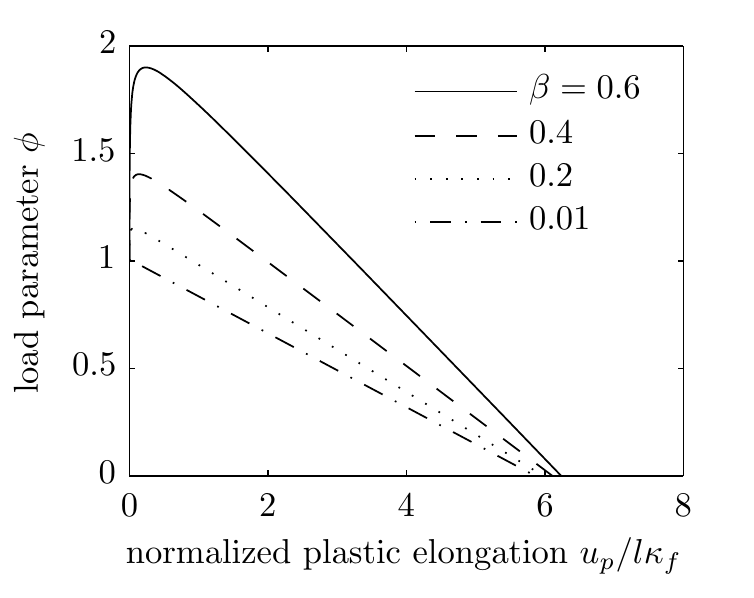}\label{SubSect:4.1:Fig:3b}}
\caption{Piecewise constant stress distribution: (a) plastic part of dimensionless load-displacement diagram for different values of dimensionless length~$\lambda_g$ assuming fixed~$\beta=0.5$, and~(b) for different values of~$\beta$ assuming fixed~$\lambda_g=0.127m_1$.}
\label{SubSect:4.1:Fig:3}
\end{figure}

The normalized plastic elongation, with the general expression presented in Eq.~\eqref{SubSect:3.2:Eq:5}, can be again evaluated analytically:
\begin{equation}
\begin{aligned}
\frac{u_p}{l\kappa_f}=&\, 2\big\{\lambda_p[1+\phi(\beta-1)]-\phi\beta\lambda_g+C_1\sin{\lambda_g}+C_3\sinh{\lambda_g}+C_5(\sin{\lambda_p}-\sin{\lambda_g})\\
&+C_6(\cos{\lambda_g}-\cos{\lambda_p})+C_7(\sinh{\lambda_p}-\sinh{\lambda_g})+C_8(\cosh{\lambda_p}-\cosh{\lambda_g})\big\}
\end{aligned}
\label{SubSect:4.1:Eq:5}
\end{equation}
Dimensionless load-plastic elongation diagrams for fixed~$\beta=0.5$ and for
different dimensionless sizes of the thin segment~$\lambda_g$ are presented in
Fig.~\ref{SubSect:4.1:Fig:3a}; the influence of~$\beta$ for fixed~$\lambda_g=0.127m_1$ with
different values of~$\beta$ is shown in Fig.~\ref{SubSect:4.1:Fig:3b}. Notice
that the obtained results resemble those presented in
Section~\ref{SubSect:3.2} for a bar with piecewise constant initial yield
stress. However, the slope of the load-displacement diagram now strongly depends on~$\lambda_g$ and~$\beta$.

Plastic admissibility condition~\eqref{SubSect:2.2:Eq:4} valid outside the plastic zone provides the inequality already presented in Eq.~\eqref{SubSect:3.2:Eq:7}, and can be simply verified in Fig.~\ref{SubSect:4.1:Fig:1}, where for~$\beta=0.5$ we require~$\phi\leq 2$.

\begin{figure}
\centering
\includegraphics[scale=1]{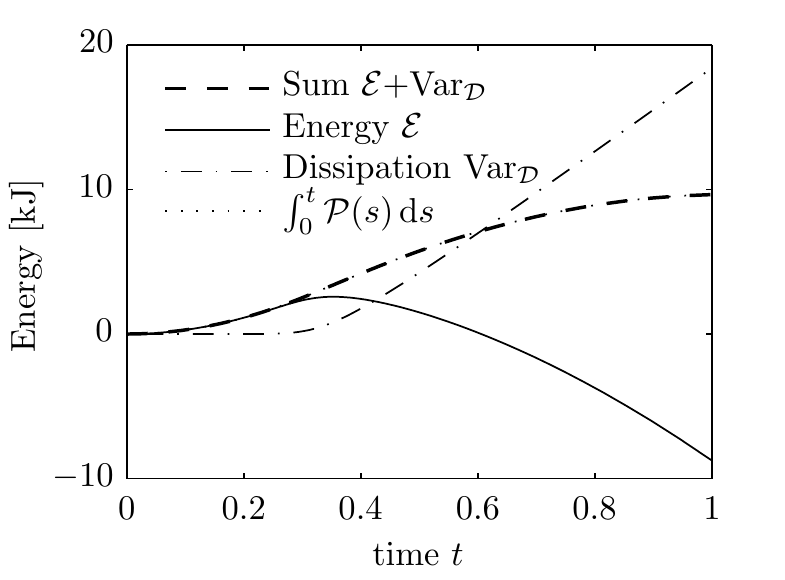}
\caption{Piecewise constant stress distribution: energy evolution for~$\beta=0.5$; see Tab.~\ref{SubSect:3.2:Tab:1} for the physical parameters and Eq.~\eqref{SubSect:3.2:Eq:8} for the loading program.}
\label{SubSect:4.1:Fig:4}
\end{figure}

Finally, the energy profiles corresponding to the loading program~\eqref{SubSect:3.2:Eq:8} are depicted in Fig.~\ref{SubSect:4.1:Fig:4}, where we can check that the solution satisfies the energy balance~\eqref{Sect:1:Eq:E} along the whole loading path. Physical constants are summarized in Tab.~\ref{SubSect:3.2:Tab:1}; the parameters reflecting the size of the thin segment were set to~$\lambda_g=0.255m_1$ and~$\beta=0.5$, and the total length of the bar was~$L=4lm_1$.
%
%-----------------------------------------------------------------------------
%	SECTION 5
%-----------------------------------------------------------------------------
%
\section{Bar With Piecewise Linear Stress Distribution}
\label{Sect:5}
Now we proceed to a bar with a continuous but not continuously differentiable stress distribution, cf. Fig.~\ref{Sect:4:Fig:1b}. The cross-sectional area corresponding to a piecewise linear stress distribution is specified in the form
\begin{equation}
A(x)=\frac{A_cl_g}{l_g-|x|}
\label{Sect:5:Eq:1}
\end{equation}
\new{where the overall length of the bar~$L=2l_g$ now represents the supremum over all possible bar lengths for which the example is meaningful; note that~$\lim_{x\rightarrow\pm l_g}A(x)=+\infty$.} As in the
previous section, $A_c$ denotes the area of the weakest cross section. Substituting~$A(x)$ into Eq.~\eqref{SubSect:2.2:Eq:2} with~$\sigma_0(x)=\sigma_r$, we obtain
\begin{equation}
l^4\left[\kappa^\mathrm{IV}(x)+\frac{2\,\sgn{x}}{l_g-|x|}\kappa'''(x)+\frac{2}{(l_g-|x|)^2}\kappa''(x)\right]-\kappa(x)=\frac{\sigma_r}{H}-\frac{\sigma_c}{Hl_g}(l_g-|x|)\mbox{ for }x\in\mathcal{I}_p\backslash\{0\}
\label{Sect:5:Eq:2}
\end{equation}
which has to be satisfied at all points inside the plastic zone with the
exception of~$x=0$, where the cross-sectional area is not differentiable.
At that point, we enforce continuity conditions of~$\kappa$, $\kappa'$, $A\kappa''$ and~$(A\kappa'')'$, recall the discussion at the end of Section~\ref{SubSect:2.1} and Tab.~\ref{SubSect:2.2:Tab:1}. Since~$A$ is continuous, the third condition actually reduces to continuity of~$\kappa''$. After conversion to the dimensionless form, in Section~\ref{SubSect:3.1}, the governing equation transforms into
\begin{equation}
\kappa_n^\mathrm{IV}(\xi)+\frac{2\,\sgn{\xi}}{\lambda_g-|\xi|}\kappa_n'''(\xi)+\frac{2}{(\lambda_g-|\xi|)^2}\kappa_n''(\xi)-\kappa_n(\xi)=\phi-1-\phi\frac{|\xi|}{\lambda_g}\mbox{ for }\xi\in(-\lambda_p,\lambda_p)\backslash\{0\}
\label{Sect:5:Eq:3}
\end{equation}
This is a fourth-order differential equation, and contrary to
Eqs.~\eqref{SubSect:3.1:Eq:2}, \eqref{SubSect:3.2:Eq:1},
and~\eqref{SubSect:4.1:Eq:1}, it has non-constant coefficients and a
non-constant right-hand side term. Although an analytical solution can be
constructed in terms of special functions---the coefficients of the homogeneous
part of equation~\eqref{Sect:5:Eq:3} fulfil the so-called Calabi-Yau condition,
cf.~\cite{CalabiYau}, Eq.~(3.4)---it is more convenient to solve it using the
MATLAB\textsuperscript{\textregistered} \texttt{bvp4c} solver, for details see~\cite{Shampine}. Again, due to symmetry conditions, it
suffices to restrict our attention to the positive part of the plastic zone, $\mathcal{I}_p^+=(0,\lambda_p)$. Then, the boundary and symmetry conditions read
\begin{equation}
\begin{aligned}
&\kappa_n(\lambda_p)=0,\kappa_n'(\lambda_p)=0,\kappa_n''(\lambda_p)=0,\mbox{ and }\\
&\kappa_n'(0)=0,\kappa_n'''(0^+)=-\kappa_n''(0)/\lambda_g
\end{aligned}
\label{Sect:5:Eq:4}
\end{equation}
The last condition is obtained from continuity of~$(A\kappa'')'$, meaning that~$(A\kappa'')'(0^-)=(A\kappa'')'(0^+)$, and can be derived when taking into account continuity of~$A$, $\kappa''$ and skew-symmetry of~$A'$, $\kappa'''$, i.e.~$A'(0^-)=-A'(0^+)$, $\kappa'''(0^-)=-\kappa'''(0^+)$. The solution is again parametrized by the length of the plastic zone~$\lambda_p$, and for each~$\lambda_p$ the corresponding~$\phi$ is determined from the solution of~\eqref{Sect:5:Eq:3} and~\eqref{Sect:5:Eq:4}.

\new{Before presenting the obtained results, let us briefly comment on the numerical solutions. Since the \texttt{bvp4c} solver is employed to provide only the continuous part of the solution on~$\mathcal{I}_p^+$, the jump condition becomes a boundary condition that is imposed directly. Moreover, the solver relies on a collocation method for iterative solution of boundary value problems with nonlinear two-point boundary conditions, and it is therefore robust with respect to possible changes of input data compared e.g.~to shooting-based strategies used in~\cite{JiZe:IJSS:2015}. For further details, we refer to~\cite{Shampine}, Section~3.}

\begin{figure}
\centering
\subfloat[]{\includegraphics[scale=1]{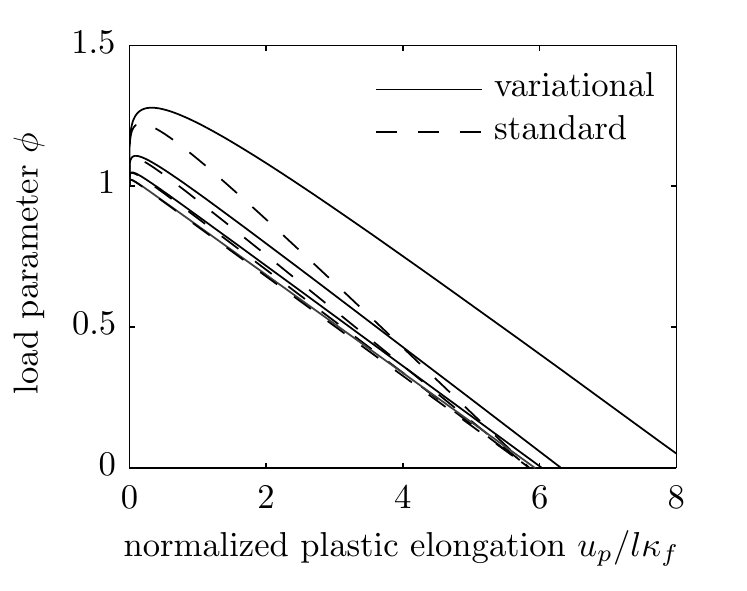}\label{Sect:5:Fig:1a}}
\subfloat[]{\includegraphics[scale=1]{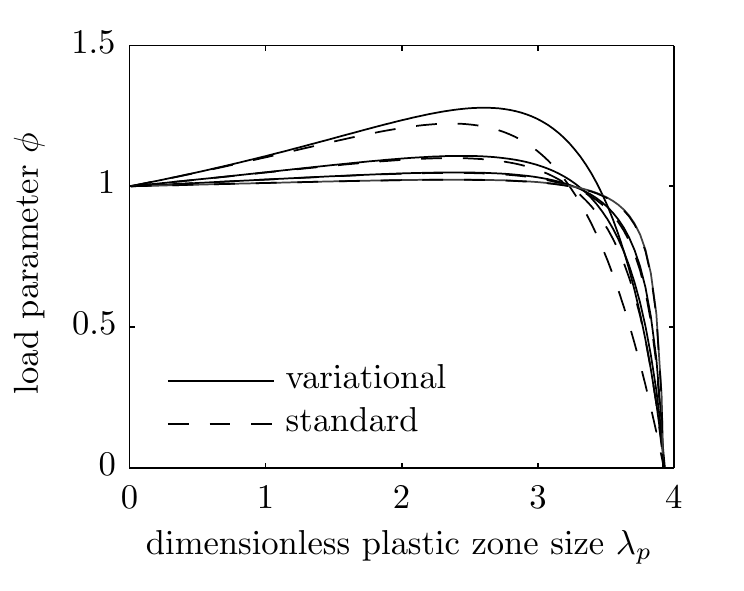}\label{Sect:5:Fig:1b}}
\caption{Piecewise linear stress distribution: (a) plastic part of load-displacement diagram, and (b) relation between load parameter and plastic zone \new{for~$\lambda_g=\{1.019,2.037,4.075,8.150\}m_1$. The maximum value of the load parameter~$\phi$ decreases with increasing~$\lambda_g$.}}
\label{Sect:5:Fig:1}
\end{figure}

The plastic part of the load-displacement diagram is depicted in
Fig.~\ref{Sect:5:Fig:1a}, where the dimensionless load parameter~$\phi$ is
plotted against the dimensionless plastic elongation~$u_p/l\kappa_f$. The
initial part of the diagram is vertical, as in the previous examples, since only
the elastic deformation evolves for~$F<F_r$. At the onset of yielding, the load
parameter first steeply increases and only later decreases. Complete failure is
attained at larger elongation in comparison with the non-variational formulation analyzed in~\cite{SoftJirZemVond}. Several values~$\lambda_g=\{1.019,2.037,4.075,8.150\}m_1$ are reported (from top to bottom), to reflect the effect of spatial variation of the sectional area; lower values of~$\lambda_g$ correspond to a stronger variation of the sectional area and lead to higher peak loads. Note that for the standard formulation, the overall elongation at failure is always the same, while for the variational approach it depends on~$\lambda_g$.

Fig.~\ref{Sect:5:Fig:1b} captures the evolution of the plastic zone size. The load parameter~$\phi$ is plotted against the dimensionless plastic zone size~$\lambda_p$, obtained again for several values of~$\lambda_g$. We can infer from the figure that the plastic zone evolves continuously and monotonically from the weakest section to its full size. The length at complete failure is almost the same as for the standard solution, plotted by dashed curves.

\begin{figure}
\centering
\subfloat[]{\includegraphics[scale=1]{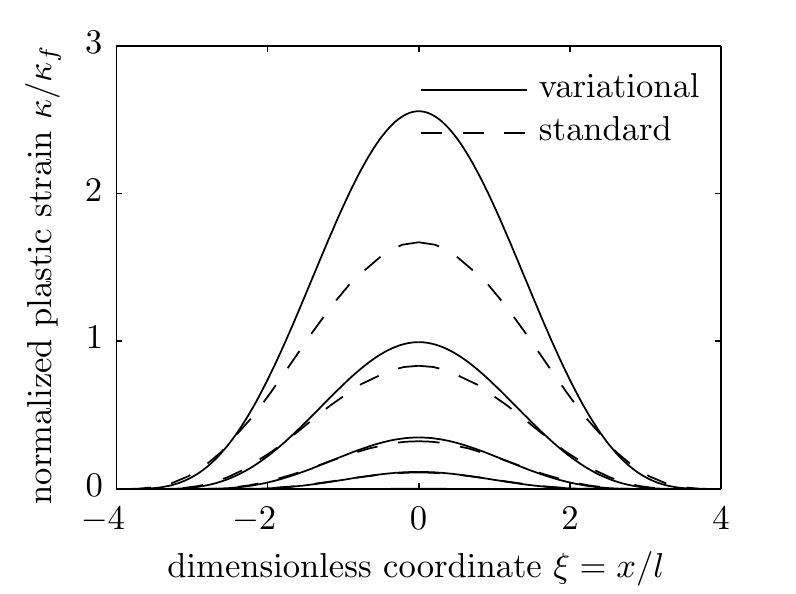}\label{Sect:5:Fig:2a}}
\subfloat[]{\includegraphics[scale=1]{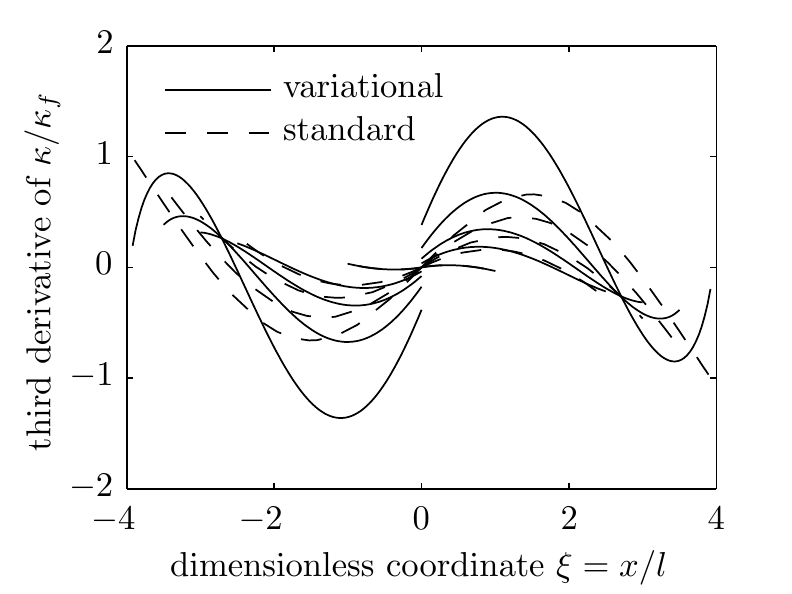}\label{Sect:5:Fig:2b}}
\caption{Piecewise linear stress distribution: (a) evolution of plastic strain profile and (b) third derivative of plastic strain for monotonically increasing plastic zone length~$\lambda_p$.}
\label{Sect:5:Fig:2}
\end{figure}

The evolution of the plastic strain profile and of its third spatial derivative is depicted in Fig.~\ref{Sect:5:Fig:2} for~$\lambda_g=1.019m_1$ and several values of~$\lambda_p=m_1\{0.255,0.637,0.764,0.891,0.998\}$. First, during the early stages of plastic evolution, the standard and variational solutions are almost the same, but at later stages, the differences grow significantly. Contrary to the standard solution, $\kappa'''$ is discontinuous for the variational solution at~$\xi=0$, where~$\llbracket\kappa'''\rrbracket_0=-2\kappa''(0)/l_g$.
\begin{figure}
\centering
\subfloat[standard]{\includegraphics[scale=1]{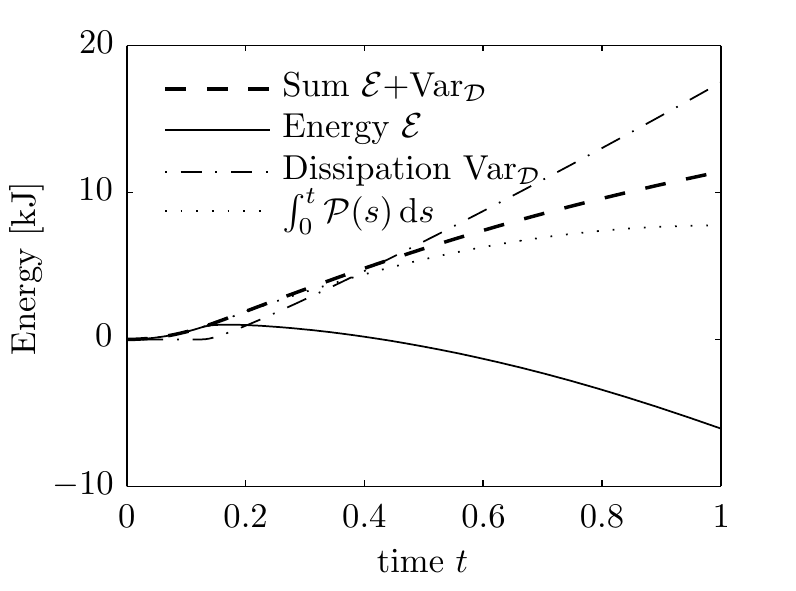}\label{Sect:5:Fig:3a}}
\subfloat[variational]{\includegraphics[scale=1]{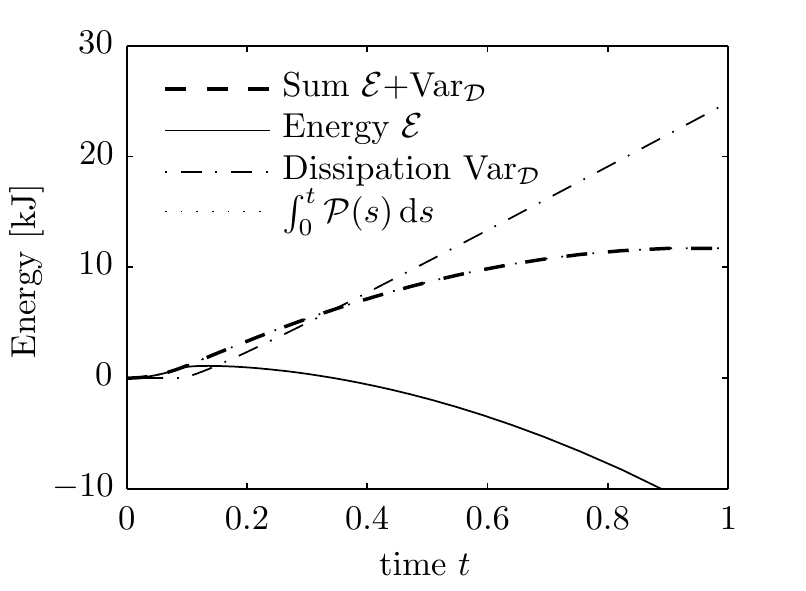}\label{Sect:5:Fig:3b}}
\caption{Piecewise linear stress distribution: energy evolutions; see Tab.~\ref{SubSect:3.2:Tab:1} for the physical parameters and Eq.~\eqref{Sect:5:Eq:6} for the loading program.}
\label{Sect:5:Fig:3}
\end{figure}

In the elastic zone~$\mathcal{I}_e$, condition~\eqref{SubSect:2.2:Eq:4} for an admissible solution reduces to
\begin{equation}
\phi\leq\frac{1}{1-\frac{\xi}{\lambda_g}}
\label{Sect:5:Eq:5}
\end{equation}
where it is sufficient to verify~$\xi=\lambda_p$. For the data in
Fig.~\ref{Sect:5:Fig:1b} we obtain
\begin{equation}
\phi\leq\frac{1}{1-\frac{1}{4}}=\frac{4}{3}
\end{equation}
and the condition is
satisfied. For~$\lambda_p\rightarrow\lambda_g$, however, the right-hand side
in~\eqref{Sect:5:Eq:5} converges to~$+\infty$ showing that plastification of
points close to physical boundary~$\partial\Omega$ would require a very strong
growth of~$\phi$.

Figure~\ref{Sect:5:Fig:3} depicts the energy balance~\eqref{Sect:1:Eq:E} for the
standard solution~(Fig.~\ref{Sect:5:Fig:3a}) and for the
energetic solution~(Fig.~\ref{Sect:5:Fig:3b}) obtained for the data presented
in Tab.~\ref{SubSect:3.2:Tab:1}. Further, we have used~$\lambda_g=1.02m_1$ and have
parametrized the localization process through
\begin{equation}
\lambda_p(t)=m_1t\mbox{ for }t\in[0,1]
\label{Sect:5:Eq:6}
\end{equation}
It is worth noting that, for the standard solution, the work done by external
forces~$\int_0^t\mathcal{P}(s)\,\mathrm{d}s$ is out of balance
with~$\mathcal{E}+\mathrm{Var}_{\mathcal{D}}$, while the variational approach delivers an energy-conserving process.
%
%-----------------------------------------------------------------------------
%	SECTION 6
%-----------------------------------------------------------------------------
%
\section{Bar With Quadratic Stress Distribution}
\label{Sect:6}
As the final example, we shall present the most regular case of quadratic stress distribution possessing continuous derivatives of an arbitrary order, see Fig.~\ref{Sect:4:Fig:1c}. The function describing the cross-sectional area then reads
\begin{equation}
A(x)=\frac{A_cl_g^2}{l_g^2-x^2}
\label{Sect:6:Eq:1}
\end{equation}
\new{where the overall length of the bar~$L=2l_g$ is, in analogy to Eq.~\eqref{Sect:5:Eq:1}, understood as the supremum over all possible bar lengths for which the example is meaningful; note that again, $\lim_{x\rightarrow\pm l_g}A(x)=+\infty$.} Upon substitution into the yield condition~\eqref{SubSect:2.2:Eq:2} with~$\sigma_0(x)=\sigma_r$, we get the governing equation
\begin{equation}
l^4\left[\kappa^\mathrm{IV}(x)+\frac{4x}{l_g^2-x^2}\kappa'''(x)+\frac{2(l_g^2+3x^2)}{(l_g^2-x^2)^2}\kappa''(x)\right]-\kappa(x)=\frac{\sigma_r}{H}-\frac{\sigma_c}{H}\left(1-\frac{x^2}{l_g^2}\right)\mbox{ for }x\in\mathcal{I}_p
\label{Sect:6:Eq:2}
\end{equation}
which can be converted into the dimensionless form
\begin{equation}
\kappa_n^\mathrm{IV}(\xi)+\frac{4\xi}{\lambda_g^2-\xi^2}\kappa_n'''(\xi)+\frac{2(\lambda_g^2+3\xi^2)}{(\lambda_g^2-\xi^2)^2}\kappa_n''(\xi)-\kappa_n(\xi)=\phi-1-\phi\frac{\xi^2}{\lambda_g^2}\mbox{ for }\xi\in(-\lambda_p,\lambda_p)
\label{Sect:6:Eq:3}
\end{equation}
As in the previous case, we will employ a numerical solver, since the governing
equation is even more complicated. Owing to symmetry requirements, the solution
will again be constructed in the positive half of the plastic
zone~$\mathcal{I}_p^+$, with boundary and symmetry conditions
\begin{equation}
\begin{aligned}
&\kappa_n(\lambda_p)=0,\kappa_n'(\lambda_p)=0,\kappa_n''(\lambda_p)=0\\
&\kappa_n'(0)=0,\kappa_n'''(0)=0
\end{aligned}
\label{Sect:6:Eq:4}
\end{equation}
In contrast to~\eqref{Sect:5:Eq:4}, the last condition is now simpler since~$A'$ is continuous.

\begin{figure}
\centering
\subfloat[]{\includegraphics[scale=1]{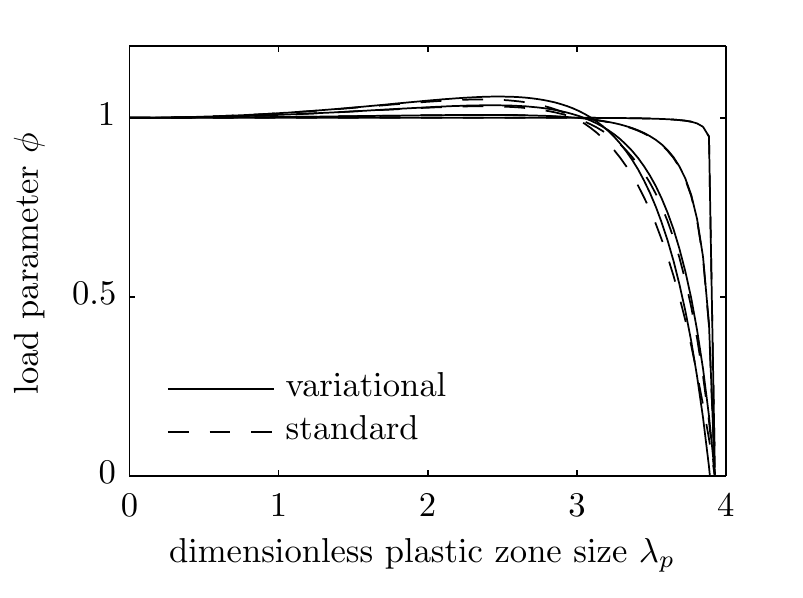}\label{Sect:6:Fig:1a}}
\subfloat[]{\includegraphics[scale=1]{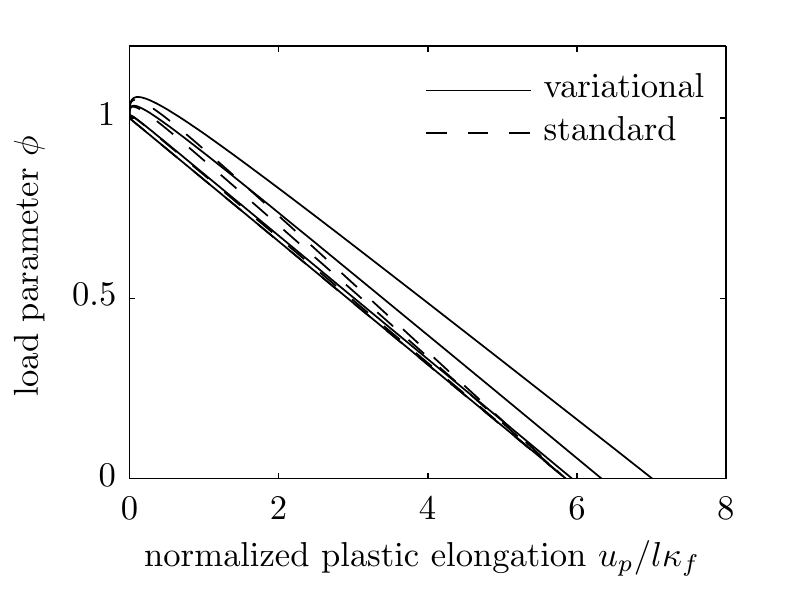}\label{Sect:6:Fig:1b}}
\caption{Quadratic stress distribution: (a) relation between load parameter and plastic zone size, (b) plastic part of load-displacement diagram \new{for~$\lambda_g=\{1.019,1.273,2.547,12.734\}m_1$. The maximum value of the load parameter~$\phi$ decreases with increasing~$\lambda_g$.}} 
\label{Sect:6:Fig:1}
\end{figure}

\begin{figure}
\centering
\subfloat[]{\includegraphics[scale=1]{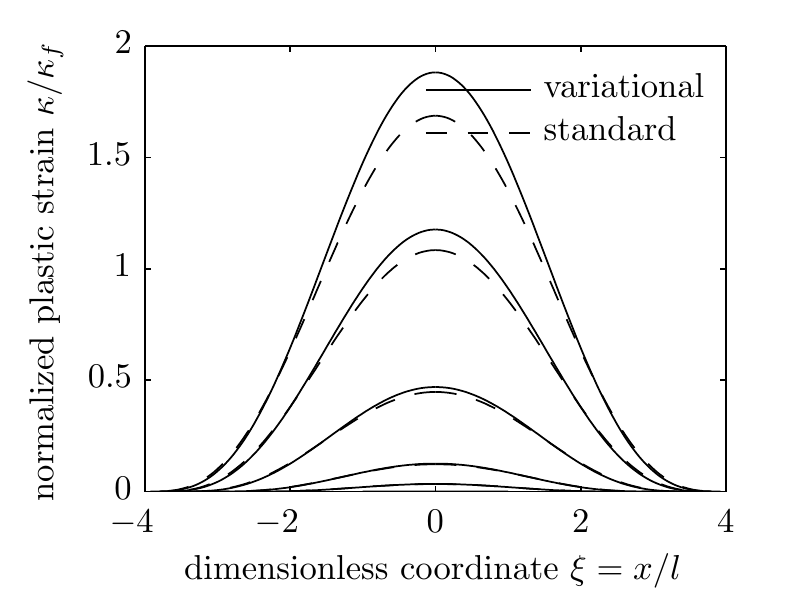}\label{Sect:6:Fig:2a}}
\subfloat[]{\includegraphics[scale=1]{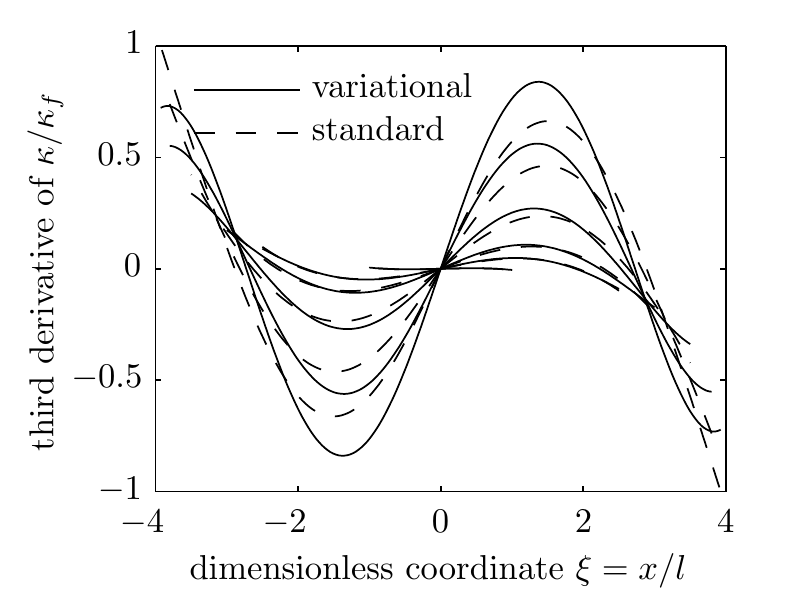}\label{Sect:6:Fig:2b}}
\caption{Quadratic stress distribution: (a) evolution of plastic strain profile and (b) third derivative of plastic strain for increasing~$\lambda_p$.}
\label{Sect:6:Fig:2}
\end{figure}

The solution has been computed for several values of~$\lambda_p$. In Fig.~\ref{Sect:6:Fig:1a} we notice that the plastic zone evolves continuously and monotonically; particular plastic strain profiles together with their third derivatives are depicted in Fig.~\ref{Sect:6:Fig:2} for~$\lambda_g=1.273m_1$ and~$\lambda_p=m_1\{0.255,0.636,0.764,0.891,0.968,0.998\}$. Comparing the results presented in Fig.~\ref{Sect:5:Fig:2} with the results in Fig.~$\ref{Sect:6:Fig:2}$, we notice that for the quadratic stress distribution the differences between the standard and variational solutions are less pronounced. Due to a higher smoothness of the solution, the load-displacement diagram presented in Fig.~\ref{Sect:6:Fig:1b} is almost linear, only with a slight hardening followed by the softening branch. Differences between plastic displacements at failure, i.e.~for~$\phi=0$, are also somewhat less distinct.

Hardening effects for the variational formulation are systematically stronger in
comparison to the standard formulation; moreover, for small values
of~$\lambda_g$, the differences are more obvious. This effect has already been
explained in Section~\ref{SubSect:2.1}, see Eq.~\eqref{SubSect:2.2:Eq:2},
Tab.~\ref{SubSect:2.2:Tab:1} and the discussion therein. Recall, nevertheless,
that the two formulations differ in two terms with higher-order derivatives of
the sectional area, neglected for the standard formulation. For decreasing
magnitudes of~$A'(x)$ and~$A''(x)$, the variational formulation approaches the
standard one; the limit case is presented in Section~\ref{SubSect:3.1}, where
the two solutions coincide. Let us note that for an infinitely differentiable
exponential stress distribution, considered in~\cite{LocAnalJir}, we would also obtain significant differences for~$\lambda_g$ small enough.

Substituting expression~\eqref{Sect:6:Eq:1} into inequality~\eqref{SubSect:2.2:Eq:4} leads to
\begin{equation}
\phi\leq\frac{1}{1-\left(\frac{\xi}{\lambda_g}\right)^2}
\label{Sect:6:Eq:5}
\end{equation}
which should hold inside~$\mathcal{I}_e$, i.e.~for all~$|\xi|\geq\lambda_p$.
A closer inspection of Fig.~\ref{Sect:6:Fig:1a} reveals that the condition is
satisfied.

\begin{figure}
\centering
\subfloat[standard]{\includegraphics[scale=1]{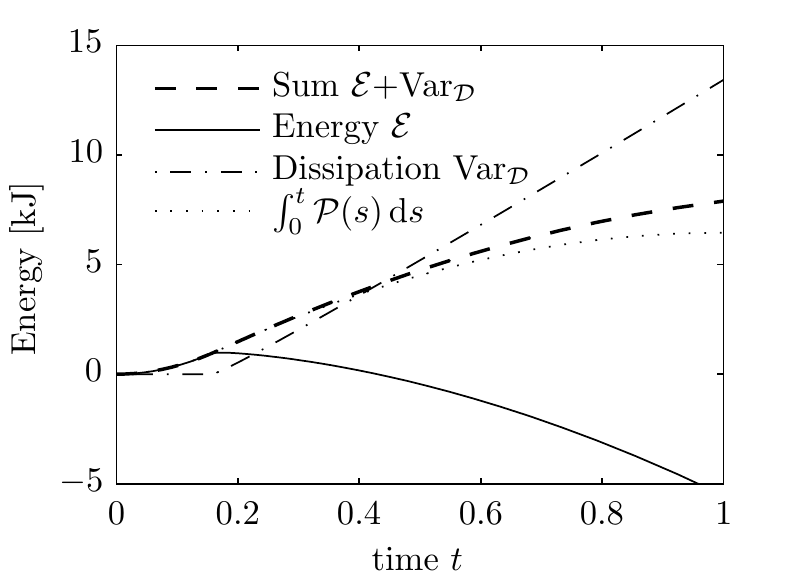}\label{Sect:6:Fig:3a}}
\subfloat[variational]{\includegraphics[scale=1]{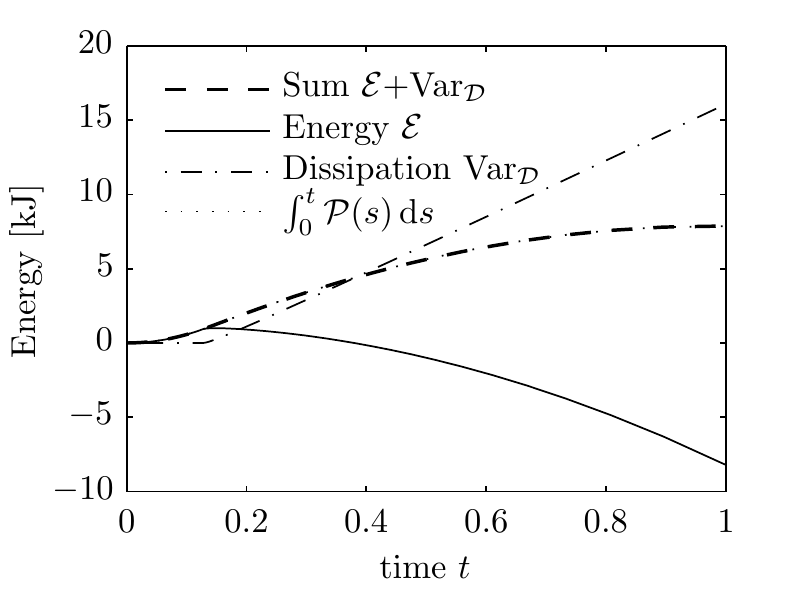}\label{Sect:6:Fig:3b}}
\caption{Quadratic stress distribution: energy evolutions; see Tab.~\ref{SubSect:3.2:Tab:1} for the physical parameters and Eq.~\eqref{Sect:5:Eq:6} for the loading program.}
\label{Sect:6:Fig:3}
\end{figure}

Energy balances for the standard and variational formulations are shown in Fig.~\ref{Sect:6:Fig:3} using the data from Tab.~\ref{SubSect:3.2:Tab:1} and loading program in Eq.~\eqref{Sect:5:Eq:6}. The total length of the bar was~$2\lambda_g$ with~$\lambda_g=1.019m_1$. Again, for the standard formulation we notice a slight violation of condition~\eqref{Sect:1:Eq:E}.
%
%-----------------------------------------------------------------------------
%	SECTION 7
%-----------------------------------------------------------------------------
%
\section{Summary and Conclusions}
\label{Sect:Summary}
We have presented one-dimensional localization analysis of a softening plasticity model
regularized by a variational formulation with a fourth-order gradient enrichment.
The main results are summarized as follows:
\begin{enumerate}
	\item Using a consistent variational approach, we have derived the description of a one-dimensional gradient plasticity model which provides not only the appropriate differential equation, representing the yield condition inside the plastic zone, but also appropriate forms of boundary and jump conditions at the elasto-plastic interface.
	
	\item On the basis of the derived conditions that follow from the variational principle, two problems with discontinuous data (a bar or layer with discontinuous yield stress and a bar with discontinuous cross-sectional area) have been investigated. These examples are amenable to analytical solution and have provided physically reasonable results.
	
\item Two additional examples have been analysed, one with continuous but not
continuously differentiable data and the other with smooth data. Numerical
solutions have been constructed and compared to the alternative non-variational formulation, demonstrating that the variationally
consistent formulation leads to higher peak loads and elongations at
structural failure.
	
\item We have also investigated the influence of various data on the evolution
of the plastic zone and on load-displacement diagrams for all four prototype
problems. It has been shown that the plastic zone monotonically expands from the
weakest section of the bar. In spite of the softening character of the material
model, the structural response exhibits first hardening after the onset of
yielding, later followed by softening. Such a behaviour is related to a gradual
expansion of the plastic zone to stronger segments of the bar.
	
\item Further, it has been demonstrated that the solution corresponding to the
variational formulation satisfies the energy balance along the evolution path of
the localization process. Contrary to that, the standard formulation exhibits
systematic lack of balance in the sense that the sum of
elastic and dissipated energies exceeds the work done by external
forces. In all cases, however, the dissipated
energy remains finite and non-zero.

	\item The variational approach is based on the non-negative first variation of the functional. Solutions corresponding to local minima of the functional have to satisfy this condition and moreover have to be stable. Hence, an analysis of the second variation providing some explicit requirements on physical constants is presented in~\ref{Sect:A} for the simplest case of a bar with perfectly uniform data.
\end{enumerate}

Let us note that although we have not employed variable elastic or plastic moduli, the
variational approach is perfectly suited to handle problems with
discontinuities in such data.

\new{The presented framework can also  be extended to higher dimensions using e.g.\ the von Mises yield function with isotropic softening. Then, the free boundary conditions for the scalar cumulative plastic strain at the elastoplastic interface are analogous to those for~$\kappa$ for sufficiently smooth fields only. However, in the multi-dimensional setting the
required regularity is difficult to establish since, for instance, the embeddings~$W^{1,2} \subset C^0$ or~$W^{2,2} \subset C^1$  no longer hold; its rigorous investigation is beyond the scope of the present work.} 
\appendix
%
%-----------------------------------------------------------------------------
%	APPENDIX A
%-----------------------------------------------------------------------------
%
\new{
\section{Second Variation and Stability Conditions}
\label{Sect:A}
The analysis presented in the main part of the paper has been based on the
condition of non-negative first variation of the energy
functional~$\Pi$. This condition is, however, only a necessary one, yet is not sufficient to ensure
that the solution is a local minimum, thus that it is energetically stable.
Therefore, in this section we discuss the behaviour of the energy
functional in the vicinity of a solution~$(u,\kappa)$
satisfying~\eqref{SubSect:2.2:Eq:1}--\eqref{SubSect:2.2:Eq:3}, drawing inspiration
from a related analysis by~\cite{LocAnalJir}.

In particular, we will investigate the second variation of the energy functional
given by Eq.~\eqref{SubSect:2.1:Eq:3}. Our objective is to show that the second variation is positive for all those nonzero admissible variations~$\delta
u$ and~$\delta\kappa$ for which the first variation~$\delta\Pi$ vanishes, 
to ensure that the solution~$(u,\kappa)$
is stable. Overall procedure will be described in six steps for better clarity. In~\ref{SubSect:A.1}, we start with the
elimination of the displacement field from the functional~$\Pi$ in order to simplify stability conditions discussed in~\ref{SubSect:A.2}, where the corresponding eigenvalue problem will be derived. In~\ref{SubSect:A.3} and~\ref{SubSect:A.4}, we discuss even and odd eigenfunctions and specify the requirements on physical and geometric parameters leading to stable responses. A discussion of larger plastic zones for a uniform bar is presented in~\ref{SubSect:A.5}. In~\ref{SubSect:A.6}, we summarize our developments and briefly comment on problems with non-uniform cross-section area.
%
%----------------------------------
%	SUBSECTION A.1
%----------------------------------
%
\subsection{Condensation of Displacement Field}
\label{SubSect:A.1}
In the first step, we can simplify the problem by eliminating the displacement field
and constructing a reduced functional that depends on the plastic strain field only. Indeed, the original
functional~\eqref{SubSect:2.1:Eq:3} can first be minimized with respect to the displacements at fixed plastic strains. 
Integrating the already derived optimality condition~\eqref{SubSect:2.2:Eq:1} and assuming vanishing body forces~$b$, we obtain
\begin{equation}\label{mj1}
u'(x) = \frac{F}{EA(x)} + \kappa(x)
\end{equation}
where~$F$ is the integration constant, physically corresponding to the axial force, cf. Eq.~\eqref{SubSect:2.2:Eq:5}. Integrating again and taking
into account the boundary conditions implied by~\eqref{Sect:1:Eq:2a}, we can express the normal force as
\begin{equation}\label{mj2}
F = K_e\left(\overline{u} - \int_{\Omega}\kappa(x)\,\mbox{d}x\right)
\end{equation}
where~$\overline{u}(t)=u_D(\partial\Omega_R,t)-u_D(\partial\Omega_L,t)$ is the prescribed total elongation, and
\begin{equation}\label{mj3}
K_e = \frac{1}{\displaystyle\int_{\Omega}\frac{\mbox{d}x}{EA(x)}} 
\end{equation}
is the elastic stiffness of the bar (reciprocal value of the elastic compliance). Based on~\eqref{mj1}--\eqref{mj2}
and on the assumption of no body forces~($b=0$),
the original functional~\eqref{SubSect:2.1:Eq:3} can be reduced to 
\begin{equation}\label{mj4}
\Pi^*(\widehat{\kappa})=\frac{K_e}{2}\left(\overline{u} - \int_{\Omega}\widehat{\kappa}\,\mbox{d}x\right)^2
+\frac{1}{2}\int_\Omega HA(\widehat{\kappa}^2-l^4\widehat{\kappa}''^2)\,\mbox{d}x+\int_\Omega A\sigma_0\widehat{\kappa}\,\mbox{d}x
\end{equation}
Note that the first term represents the elastically stored energy, expressed as a functional dependent on the plastic
strain only. The expression in the parentheses is the elastic elongation, written as the difference between the total
elongation~$\overline{u}$ and the plastic part of elongation~$u_p$ which was introduced in~\eqref{SubSect:3.2:Eq:5}.
%
%----------------------------------
%	SUBSECTION A.2
%----------------------------------
%
\subsection{Second Variation}
\label{SubSect:A.2}
The first variation of the reduced functional~$\Pi^*$ is given by
\begin{equation}\label{mj5}
\delta\Pi^*({\kappa};\delta\kappa)=-K_e\left(\overline{u} - \int_{\Omega}{\kappa}\,\mbox{d}x\right)\int_{\Omega}\delta\kappa\,\mbox{d}x
+\int_\Omega HA(\kappa\,\delta\kappa-l^4\kappa''\delta\kappa'')\,\mbox{d}x+\int_\Omega A\sigma_0\delta\kappa\,\mbox{d}x
\end{equation}
which exactly corresponds to~\eqref{SubSect:2.1:Eq:4} with~$u'$ and~$\delta u'$ expressed according to~\eqref{mj1}--\eqref{mj2} and~$b$ set to zero. The second variation of the reduced functional~$\Pi^*$ is given by
\begin{equation}\label{mj11}
\delta^2\Pi^*(\delta\kappa)=K_e\left(\int_{\Omega}\delta\kappa\,\mbox{d}x\right)^2
+\int_\Omega HA(\delta\kappa^2-l^4\delta\kappa''^2)\,\mbox{d}x
\end{equation}
and is independent of~$\kappa$ because~$\Pi^*$ is quadratic.

We would now like to prove that the quadratic form~$\delta^2\Pi^*$ is positive definite in the space
of all those admissible variations~$\delta\kappa$ for which~$\delta\Pi^*(\kappa;\delta\kappa)=0$.
Note that~$\kappa$ is not arbitrary but represents the solution of the problem described by conditions~\eqref{SubSect:2.2:Eq:3a}--\eqref{SubSect:2.2:Eq:3c},
which were derived from the requirement~$\delta\Pi(u,\kappa;\delta u,\delta\kappa)\ge 0$ for all admissible variations~$\delta u$ and~$\delta\kappa$ (and which could alternatively be derived, in a slightly different 
format, from the condition~$\delta\Pi^*(\kappa;\delta\kappa)\ge 0$ for all admissible variations~$\delta\kappa$).
Integrating the term with~$\delta\kappa''$ in~\eqref{mj5} twice by parts and exploiting the properties of function~$\kappa$
(in particular, the fact that~$\kappa=0$ in~$\mathcal{I}_e$), we can rewrite the first variation of~$\Pi^*$ as
\begin{eqnarray}\nonumber
\delta\Pi^*({\kappa};\delta\kappa)&=&
\int_{\mathcal{I}_p} \left[ HA\kappa-(HAl^4\kappa'')''+A\sigma_0-K_e\left(\overline{u} - \int_{\mathcal{I}_p}{\kappa}\,\mbox{d}x\right)\right]\delta\kappa\,\mbox{d}x\\
\nonumber
&&+\int_{\mathcal{I}_e} \left[A\sigma_0-K_e\left(\overline{u} - \int_{\mathcal{I}_p}{\kappa}\,\mbox{d}x\right)\right]\delta\kappa\,\mbox{d}x\\
&&-\sum_{\partial\mathcal{I}_{ep}}HAl^4\kappa''n\delta\kappa'+\sum_{\partial\mathcal{I}_{ep}}(HAl^4\kappa'')'n\delta\kappa
\label{mj13}
\end{eqnarray}
where~$n$ denotes the outer normal to~$\mathcal{I}_p$.
Taking into account that~$\kappa$ satisfies the yield condition~\eqref{SubSect:2.2:Eq:2} in which the right-hand side corresponds to the normal force~$F$,
which is in turn equal to the expression given in~\eqref{mj2}, we can show that the expression in square brackets in the
first integral in~\eqref{mj13} vanishes. The first sum in~\eqref{mj13} also vanishes because~$\kappa''=0$ on~$\partial\mathcal{I}_{ep}$.
Consequently, \eqref{mj13} reduces to
\begin{equation}\label{mj14}
\delta\Pi^*({\kappa};\delta\kappa)=\int_{\mathcal{I}_e} \left[A\sigma_0-K_e\left(\overline{u} - \int_{\mathcal{I}_p}{\kappa}\,\mbox{d}x\right)\right]\delta\kappa\,\mbox{d}x+\sum_{\partial\mathcal{I}_{ep}}(HAl^4\kappa'')'n\delta\kappa
\end{equation}

Let us also recall that (i)~the variation~$\delta\kappa$ is nonnegative
in~$\mathcal{I}_e\cup \partial\mathcal{I}_{ep}$; (ii)~the expression in the square brackets in~\eqref{mj14} is nonnegative in~$\mathcal{I}_e$, see condition~\eqref{SubSect:2.2:Eq:4};
and (iii)~the product~$(HAl^4\kappa'')'n$ is nonnegative on~$\partial\mathcal{I}_{ep}$, see the last line in Table~\ref{SubSect:2.2:Tab:1}.
To proceed further, we need to assume that the expression in the square brackets in~\eqref{mj14} is strictly positive (not just nonnegative) in~$\mathcal{I}_e$.
This assumption means that the normal force in the entire elastic domain is below the yield limit, which is true for all the localized solutions presented in this paper. Consequently, for those admissible variations~$\delta\kappa$ that are not identically zero in~$\mathcal{I}_e$,
the integral in~\eqref{mj14} is positive and thus the first variation~$\delta\Pi^*$ is positive, because the second term on the right-hand side of~\eqref{mj14} is always nonnegative.  

Based on the foregoing analysis of the first variation, we can see that
positive definiteness of the second variation~$\delta^2\Pi^*$ needs to be proven for those variations~$\delta\kappa$
that vanish in~$\mathcal{I}_e$, and the integration domains in~\eqref{mj11} can be restricted to~$\mathcal{I}_p$.
To preserve continuous differentiability of~$\delta\kappa$ in~$\Omega$, 
its value and first derivative on~$\partial\mathcal{I}_{ep}\equiv\partial\mathcal{I}_p$
must be zero. To simplify notation, we will use from now on~$v$ instead of~$\delta\kappa$. The stability condition is then satisfied if there exists some~$\alpha>0$ such that
\begin{equation}\label{mj21}
K_e\left(\int_{\mathcal{I}_p}v\,\mbox{d}x\right)^2
+\int_{\mathcal{I}_p} HA(v^2-l^4v''^2)\,\mbox{d}x \geq \alpha \Vert v \Vert^2 \hskip 5mm \mbox{for all } v\in\mathcal{V}
\end{equation}
where
\begin{equation}
\mathcal{V}= \widetilde{W}^{2,2}_0(\mathcal{I}_p)=\Big\{v\in W^{2,2}(\mathcal{I}_p)\,|\,\left.v\right|_{\partial\mathcal{I}_p}=0,\left.v'\right|_{\partial\mathcal{I}_p}=0\Big\}
\end{equation}
and where~$\Vert v \Vert$ denotes standard $L^2(\mathcal{I}_p)$-norm of a function~$v$.

Since~$K_e>0$ and~$HA<0$, it is instructive to rewrite inequality~\eqref{mj21} as
\begin{equation}\label{mj22}
K_e\left(\int_{\mathcal{I}_p}v\,\mbox{d}x\right)^2
-\int_{\mathcal{I}_p} HAl^4v''^2\,\mbox{d}x \geq -\int_{\mathcal{I}_p} HAv^2\,\mbox{d}x + \alpha \Vert v \Vert^2
\end{equation}
or, in a more abstract form, as
\begin{equation}\label{mj23}
(v,v)_A \geq (v,v)_B + \alpha \Vert v \Vert^2
\end{equation}
where~$(v,v)_A$ and~$(v,v)_B$ are quadratic forms corresponding to symmetric bilinear forms
\begin{eqnarray}\label{mj24}
(v,w)_A &=& K_e\int_{\mathcal{I}_p}v\,\mbox{d}x\int_{\mathcal{I}_p}w\,\mbox{d}x
-\int_{\mathcal{I}_p} HAl^4v''w''\,\mbox{d}x \\
(v,w)_B &= & -\int_{\mathcal{I}_p} HAvw\,\mbox{d}x
\end{eqnarray}
Since form~$(v,w)_B$ is positive definite and form~$(v,w)_A$ is at least positive semidefinite
(in fact it is positive definite), condition~\eqref{mj23} can be reformulated as
\begin{equation}\label{mj25}
\lambda_{\min}\equiv \inf_{v\in\mathcal{V}\setminus\{0\}}\frac{(v,v)_A}{(v,v)_B} > 1
\end{equation}
Indeed, if~\eqref{mj25} is satisfied, then we have (for all~$v\in\mathcal{V}$)
\begin{equation}
(v,v)_A \geq \lambda_{\min}(v,v)_B = (v,v)_B + (\lambda_{\min}-1)(v,v)_B \geq (v,v)_B + (\lambda_{\min}-1)(-HA)_{\min}  \Vert v \Vert^2
\end{equation}
and thus inequality~\eqref{mj21} is satisfied with~$\alpha=(\lambda_{\min}-1)(-HA)_{\min}$, where~$(-HA)_{\min}>0$ is a lower bound on~$-HA$ over~$\mathcal{I}_p$. 

The fraction in~\eqref{mj25} is the well-known Rayleigh quotient, and~$\lambda_{\min}$ is the minimum eigenvalue obtained from 
the eigenvalue problem 
\begin{equation}\label{mj26}
(v,w)_A -\lambda (v,w)_B = 0 \hskip 5mm \mbox{for all } w\in \mathcal{V}
\end{equation}
Note that the rigorous treatment of~\eqref{mj25}, its transition to~\eqref{mj26}, and relation to the Poincar\'{e} inequality can be found in~\cite{Vejch}.

To finish the stability analysis, it is necessary to find the eigenvalues~$\lambda$ for which~\eqref{mj26} 
has a nontrivial solution~$v$ and check that they are greater than~1. In general, this would need to be done
numerically. However, it is instructive to construct an analytical solution
for the simplest case of a uniform bar, with $EA=$~const., $HA=$~const., and~$K_e=EA/L$. 
The eigenvalue problem~\eqref{mj26} then reads
\begin{equation}\label{mj27}
\frac{EA}{L}\int_{\mathcal{I}_p}v\,\mbox{d}x\int_{\mathcal{I}_p}w\,\mbox{d}x-HAl^4\int_{\mathcal{I}_p} v''w''\,\mbox{d}x +\lambda HA\int_{\mathcal{I}_p} vw\,\mbox{d}x = 0 \hskip 5mm \mbox{for all } w\in \mathcal{V}
\end{equation}
Integrating by parts and exploiting the boundary conditions~$w=0$ and~$w'=0$ on~$\partial\mathcal{I}_p$,
we can construct the corresponding eigenvalue problem in terms of differential (and in this case also integral) operators:
\begin{equation}\label{mj28}
\frac{E}{L}\int_{\mathcal{I}_p}v\,\mbox{d}x
-Hl^4 v^{IV} +\lambda H v = 0 
\end{equation}
This is an integro-differential equation, which can be for convenience converted into the set of two equations,
\begin{eqnarray}\label{mj29}
l^4 v^{IV} -\omega^4  v &=& C_0 \\
\frac{E}{HL}\int_{\mathcal{I}_p}v\,\mbox{d}x &=& C_0 
\label{mj29b}
\end{eqnarray}
Here, $\omega \equiv \lambda^{1/4}$, just to simplify the subsequent derivations, and~$C_0$ is an auxiliary unknown. 

The advantage of the transformation of~\eqref{mj28} into~\eqref{mj29}--\eqref{mj29b} is that~\eqref{mj29} is a standard linear differential equation
with constant coefficients and constant right-hand side, and its general solution is easily expressed as
\begin{equation}\label{mj30}
v(x) = -\frac{C_0}{\omega^4} + C_1 \cos \frac{\omega x}{l}  + C_2 \sin \frac{\omega x}{l}  + C_3 \cosh \frac{\omega x}{l}  + C_4 \sinh \frac{\omega x}{l} 
\end{equation}
Recall that the plastic zone for the most localized solution in a uniform bar corresponds to the interval~$\mathcal{I}_p=(-m_1 l,m_1 l)$ where~$m_1 \doteq 3.9266$
is the smallest positive solution of the equation~$\tan x = \tanh x$. The solution~$v\in\mathcal{V}$ must satisfy boundary conditions~$v(\pm m_1l)=0$ and~$v'(\pm m_1l)=0$, which provide four equations for five unknown constants, $C_0$ to~$C_4$. The fifth
equation is obtained from~\eqref{mj29b}. By simple row manipulations, the resulting set of five homogeneous linear equations
can be decoupled into two independent sets, which read
\begin{eqnarray}\label{mj32}
-\omega^{-4} C_0 + C_1\cos\omega m_1 + C_3\cosh\omega m_1 &=& 0\\
-C_1\sin\omega m_1 + C_3\sinh\omega m_1 &=& 0 \\
-\left(\frac{m_1}{\omega^3}+\frac{HL\omega}{2El}\right)C_0+C_1\sin\omega m_1 + C_3\sinh\omega m_1 &=& 0 
\label{mj32c}
\end{eqnarray}
and
\begin{eqnarray}\label{mj31}
C_2\sin\omega m_1 + C_4\sinh\omega m_1 &=& 0 \\
C_2\cos\omega m_1 + C_4\cosh\omega m_1 &=& 0
\label{mj31b}
\end{eqnarray}
%
%----------------------------------
%	SUBSECTION A.3
%----------------------------------
%
\subsection{Even Eigenfunctions}
\label{SubSect:A.3}
\begin{figure}
	\centering
	\subfloat[]{\includegraphics[scale=1]{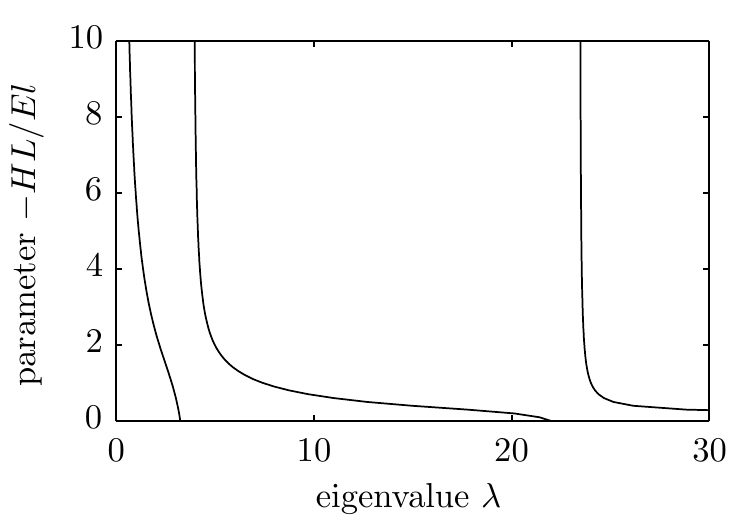}\label{mjfig2a}}\hspace{0.5em}
	\subfloat[]{\includegraphics[scale=1]{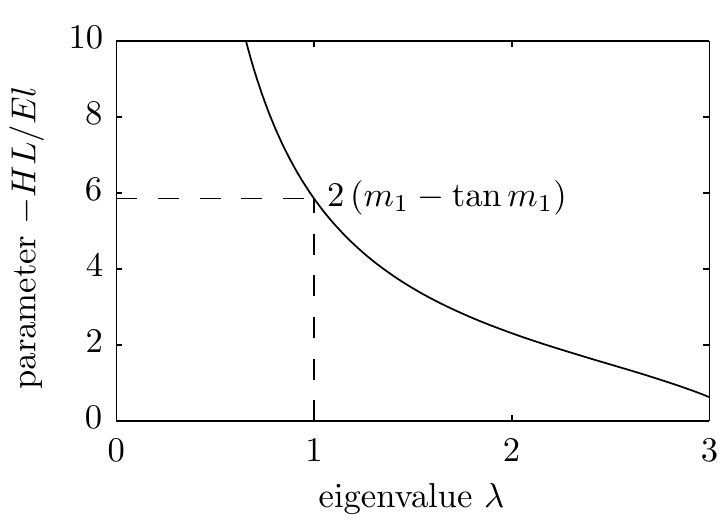}\label{mjfig2b}}
	\caption{Relation between dimensionless parameter~$-HL/El$ and eigenvalues of the first kind.}
	\label{mjfig2}
\end{figure}
Let us first examine~\eqref{mj32}--\eqref{mj32c}, having a nontrivial solution if and only if
\begin{equation}\label{mj36}
2\omega^{-4}\sin\omega m_1\sinh\omega m_1 -\left(\frac{m_1}{\omega^3}+\frac{HL\omega}{2El}\right)\left(\cos\omega m_1\sinh\omega m_1+\sin\omega m_1\cosh\omega m_1\right) = 0
\end{equation}
which is equivalent to
\begin{equation}\label{mj37}
\frac{2}{\omega^5}\left(m_1\omega-\frac{2\sin\omega m_1\sinh\omega m_1}{\cos\omega m_1\sinh\omega m_1+\sin\omega m_1\cosh\omega m_1}\right) = -\frac{HL}{El}
\end{equation}
Positive solutions of this transcendental equation, $\omega^{(1)}_i$, $i=1,2,\ldots$, correspond to eigenvalues~$\lambda^{(1)}_i=\left(\omega^{(1)}_i\right)^4$, $i=1,2,\ldots$.
Superscript~$^{(1)}$ means that we are dealing with solutions of the ``first kind'', resulting from 
singularity of~\eqref{mj32}--\eqref{mj32c}, with nonzero constants~$C_0$, $C_1$ and~$C_3$ and with~$C_2=C_4=0$. Consequently, the corresponding eigenfunctions are even. 
The eigenvalues cannot be expressed  analytically, but one can characterize them graphically, by plotting
the left-hand side of~\eqref{mj37} as a function of~$\lambda\equiv\omega^4$, as shown in Fig.~\ref{mjfig2}. The vertical
axis corresponds to the dimensionless ratio~$-HL/El$. For a given bar, this ratio is fixed and the intersections
of the corresponding horizontal line with the graph provide the eigenvalues for the considered case. 
\begin{figure}
	\centering
	\subfloat[]{\includegraphics[scale=1]{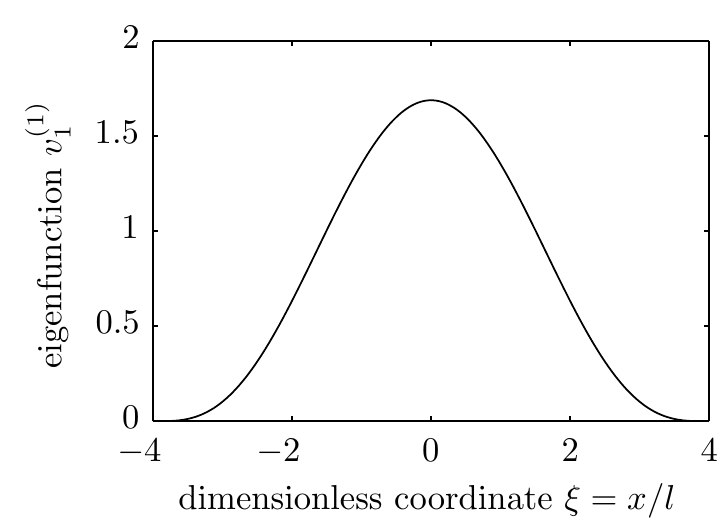}\label{mjfig1a}}\hspace{0.5em}
	\subfloat[]{\includegraphics[scale=1]{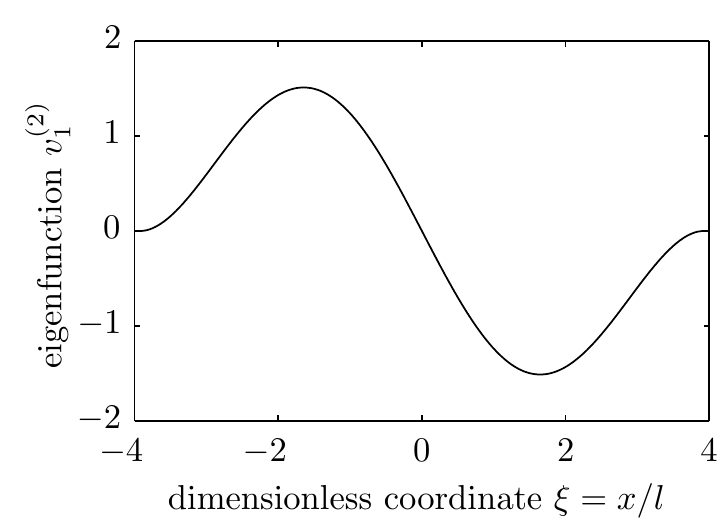}\label{mjfig1b}}
	\caption{Eigenfunctions corresponding to the smallest eigenvalue (a)~of the first kind, (b)~of the second kind.}
	\label{mjfig1}
\end{figure}

From Fig.~\ref{mjfig2} it is clear that 
if~$-HL/El$ is at or above a certain critical value, the smallest eigenvalue of the first kind 
is smaller than or equal to~1 and the stability condition is violated. The critical value is obtained simply by evaluating the
left-hand side of~\eqref{mj37} for~$\omega=1$, and turns out to be equal to~$2\,(m_1-\tan m_1)\doteq 5.8548$. 
Thus the resulting stability condition reads 
\begin{equation}\label{mj38}
-\frac{HL}{El} < 2\,(m_1-\tan m_1)
\end{equation}
This is a constraint that involves the ratio between the absolute value of the softening modulus and the elastic modulus,
$-H/E$, and the ratio between the bar length and the material characteristic length, $L/l$.
Stability (under displacement control, i.e., with~$\overline{u}$ prescribed) is compromised for long bars
(large~$L/l$) made of materials with pronounced softening (large~$-H/E$). The eigenfunction 
\begin{equation}\label{mj39}
v^{(1)}_1 = 1-\frac{\cos\displaystyle\frac{x}{l}}{2\cos m_1}-\frac{\cosh\displaystyle\frac{x}{l}}{2\cosh m_1}, \hskip 5mm x\in\mathcal{I}_p\equiv (-m_1l,m_1l)
\end{equation}
corresponding to the smallest eigenvalue of the first kind in the critical case when~$\lambda^{(1)}_1=1$
(i.e., when~$-{HL}/{El} = 2\,(m_1-\tan m_1)$) is plotted in Fig.~\ref{mjfig1a}. 
Up to
an arbitrary scalar multiplication factor, it is identical  with the actual localized solution~$\kappa(x)$ described in normalized form
by equations~\eqref{SubSect:3.1:Eq:2} and~\eqref{SubSect:3.1:Eq:7}. 
The critical case corresponds to a snapback point of the load-displacement diagram, i.e., to a point
at which the tangent becomes vertical and the amplitude of the plastic strain profile can grow at constant
total elongation of the bar while the equilibrium equation and the yield condition remain satisfied.
%
%----------------------------------
%	SUBSECTION A.4
%----------------------------------
%
\subsection{Odd Eigenfunctions}
\label{SubSect:A.4}
We still need to examine another set of eigenvalues, referred to as eigenvalues of the ``second kind''.
Equations~\eqref{mj31}--\eqref{mj31b} have a nontrivial solution if and only if
\begin{equation}\label{mj33}
\sin\omega m_1\cosh\omega m_1 - \cos\omega m_1\sinh\omega m_1 = 0
\end{equation}
which is satisfied for
\begin{equation}\label{mj34}
\omega^{(2)}_i = \frac{m_i}{m_1}, \hskip 5mm i=1,2,\ldots
\end{equation}
The corresponding eigenvalues and eigenfunctions of problem~\eqref{mj28} are
\begin{align}
\lambda^{(2)}_i &=\left(\omega^{(2)}_i\right)^4 = \left(\frac{m_i}{m_1}\right)^4, \hskip 5mm i=1,2,\ldots \label{mj35}\\
v^{(2)}_i &= \frac{\sin\displaystyle\frac{m_ix}{m_1l}}{\sin m_i} - \frac{\sinh\displaystyle\frac{m_ix}{m_1l}}{\sinh m_i}, \hskip 5mm x\in\mathcal{I}_p\equiv (-m_1l,m_1l) \label{mj35a}
\end{align}
The smallest eigenvalue of the second kind, 
$\lambda^{(2)}_1$, is equal to~1, while all the others are greater than~1. All eigenfunctions of the second
kind are odd, and the eigenfunction~$v^{(2)}_1$ corresponding to the smallest eigenvalue is plotted
in Fig.~\ref{mjfig1b}.
%
%----------------------------------
%	SUBSECTION A.5
%----------------------------------
%
\subsection{Note on Larger Plastic Zones}
\label{SubSect:A.5}
The analysis presented so far for a uniform bar has referred exclusively to the
localized solutions with the smallest possible plastic zone of size~$L_p=2m_1l$.   
The original set of conditions~\eqref{SubSect:2.2:Eq:3a}--\eqref{SubSect:2.2:Eq:3c}, from which the plastic strain distribution~$\kappa(x)$ was determined, may admit other solutions, with larger plastic zones.
Indeed, the plastic zone size follows from equation~\eqref{SubSect:3.1:Eq:6}, which has positive solutions~$\lambda_p = m_i$, $i=1,2,\ldots$, corresponding to plastic zone sizes~$L_p = 2m_il$, $i=1,2,\ldots$. In Section~\ref{SubSect:3.1} we decided to focus on the most
localized solution with~$i=1$. If a uniform bar 
is long enough,  there exist other solutions, corresponding to~$i\ge 2$. 
An example for~$i=2$ is plotted in Fig.~\ref{mjfig3a}.
Such solutions satisfy conditions that follow from non-negativeness of the first
variation, but they violate the stability condition and thus cannot physically occur.
To see that, consider the case of~$i=2$, i.e., $\mathcal{I}_p= (-m_2l,m_2l)$
where~$m_2\doteq 7.0686$. Condition~\eqref{mj33} is then replaced by
\begin{equation}\label{mj43}
\sin\omega m_2\cosh\omega m_2 - \cos\omega m_2\sinh\omega m_2 = 0
\end{equation}
Eigenvalues of the second kind are given by
\begin{equation}\label{mj44}
\lambda^{(2)}_i =\left(\omega^{(2)}_i\right)^4 = \left(\frac{m_i}{m_2}\right)^4, \hskip 5mm i=1,2,\ldots
\end{equation}
and the minimum eigenvalue, $\lambda^{(2)}_1 = (m_1/m_2)^4$, is smaller than~1.
This indicates that condition~\eqref{mj25} is violated and stability is lost. Fig.~\ref{mjfig3b}
shows the eigenfunction~$v^{(2)}_1$ corresponding to the minimum eigenvalue.
\begin{figure}
	\centering
	% \psfragfig{matfrag/mfig3a}
	% \psfragfig{matfrag/mfig3b}
	\subfloat[]{\includegraphics[scale=1]{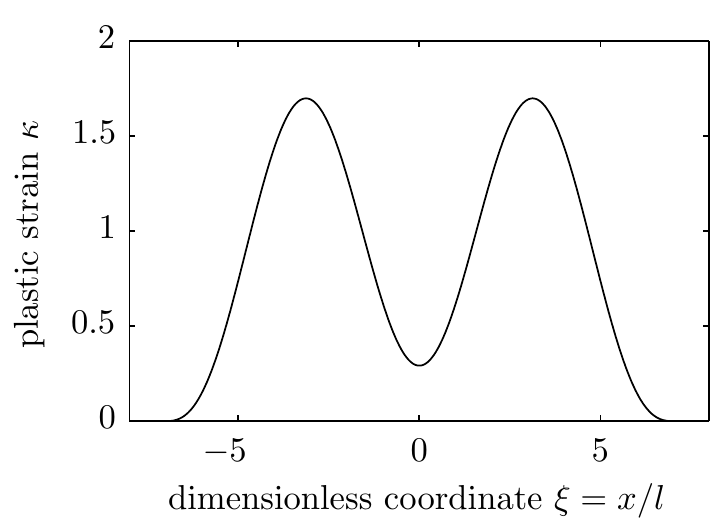}\label{mjfig3a}}\hspace{0.5em}
	\subfloat[]{\includegraphics[scale=1]{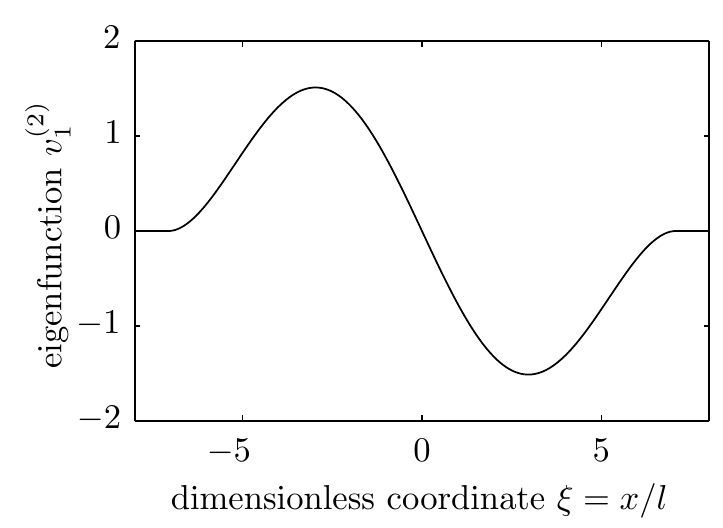}\label{mjfig3b}}
	\caption{Eigenfunctions corresponding to the smallest eigenvalue (a)~of the first kind, (b)~of the second kind.}
	\label{mjfig3}
\end{figure}
%
%----------------------------------
%	SUBSECTION A.6
%----------------------------------
%
\subsection{Summary}
\label{SubSect:A.6}
In summary, for a uniform bar (or sufficiently long uniform weakest segment of a bar) we have found that 
eigenvalues of the first kind are greater than~1 if the material constants and bar length satisfy a certain
constraint, but the smallest eigenvalue of the second kind is always equal to~1. Strictly speaking,
this would mean that stability cannot be guaranteed, because inequality~\eqref{mj21} cannot be satisfied
with any positive~$\alpha$ but only with~$\alpha=0$. The second variation is then non-negative but not
positive definite. This is related to the special nature of the localization problem for a uniform bar,
which actually admits infinitely many localized solutions characterized by the same energy level.   
Indeed, the precise position of the localized zone in a perfectly uniform bar remains undetermined,
and the considered solution (centered in the middle of the bar) can be arbitrarily shifted without
modifying the total energy. In a real bar, the position of the localized plastic zone would be affected by 
imperfections of the geometry and material data (sectional area, initial yield stress, etc.)
For an ideal, perfectly uniform bar, the localized plastic zone can form anywhere and the corresponding
solutions are neutrally stable. It is not by chance that the eigenfunction~$v^{(2)}_1$ corresponding
to eigenvalue~$\lambda^{(2)}_1=1$ is actually a scalar multiple of the spatial derivative of function~$\kappa$
which describes the 
plastic strain distribution. Adding an infinitely small multiple of~$\kappa'$ to~$\kappa$ corresponds to
an infinitesimal horizontal shift of the localized plastic strain profile. Note that adding a finite multiple
of~$\kappa'$ would result into violation of the constraint~$\kappa\ge 0$ near one boundary of the plastic zone
and is thus inadmissible. Shifted plastic strain profiles form a parametric family of functions at the same
energy level which does not correspond to a linear manifold in the functional space. Convex linear combinations
of two selected members of this family correspond to higher energy levels and if the difference of two such
functions is considered as a variation~$\delta\kappa$, the corresponding first variation~$\delta\Pi^*$ is positive.
This is in agreement with our conclusion that variations which do not vanish everywhere in the elastic zone
lead to positive~$\delta\Pi^*$ and thus do not need to be considered in the stability analysis based on
the second variation.

Before closing \ref{Sect:A}, let us demonstrate the presented approach for a bar with a variable cross-sectional area. In particular, we verify~$\lambda_1^{(2)} > 1$ by numerically solving the eigenvalue problem~\eqref{mj26} for a family of cross-sectional areas
\begin{equation}
\begin{aligned}
A_l(x) &= 1 + \frac{a|x|}{1-a|x|}\\
A_q(x) &= 1 + \frac{(ax)^2}{1-(ax)^2}
\end{aligned}
\quad\quad\mbox{for } x \in \mathcal{I}_p \equiv (-m_1l,m_1l)
\label{mj45}
\end{equation}
where~$a\in[0,\frac{1}{m_1l})$, cf.~also Eqs.~\eqref{Sect:5:Eq:1} and~\eqref{Sect:6:Eq:1}, corresponding to piecewise linear and quadratic stress distributions. The obtained results are depicted in Fig.~\ref{mjfig4} where we verify that for~$a \rightarrow 0$ we get $\lambda_1^{(2)} \rightarrow 1$, and that $\lambda_1^{(2)} > 1$ for~$a \neq 0$, as expected.
\begin{figure}
	\centering
	% \psfragfig{matfrag/eigSecond}
	\includegraphics[scale=1]{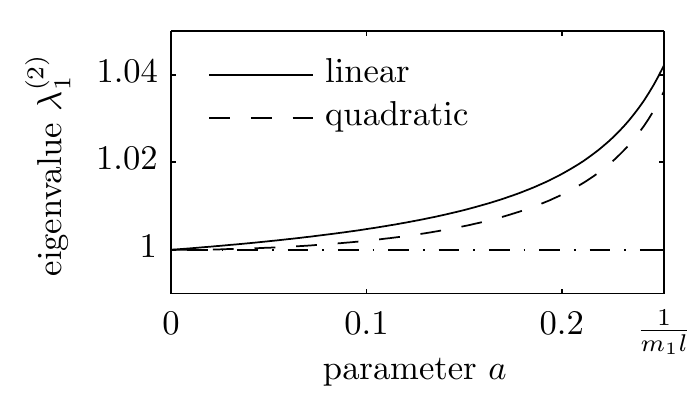}
	\caption{Eigenvalue of the second kind, $\lambda_1^{(2)}$, as a function of~$a\in[0,\frac{1}{m_1l})$ for cross-sectional areas corresponding to piecewise linear and quadratic stress distributions, see Eq.~\eqref{mj45}.}
	\label{mjfig4}
\end{figure}
This numerical result confirms that nonuniformity of the cross-section area~$A(x)$ has a stabilizing effect.
}
%
%-----------------------------------------------------------------------------
%	APPENDIX B
%-----------------------------------------------------------------------------
%
\section{Stability of the Homogenenous Boundary
Condition~\texorpdfstring{$\kappa''=0$}{k'' = 0}}
\label{Sect:B}
In order to verify that~$\llbracket HAl^4\kappa''\rrbracket=0$ is the
correct optimality condition at the elasto-plastic interface, we construct a family
of solutions without taking this condition into account, and then prove that the
energy-minizing state coincides with the solutions presented above.

For simplicity, we consider a uniform bar described
by~\eqref{SubSect:3.1:Eq:2} with a symmetric general solution presented
in~\eqref{SubSect:3.1:Eq:3}, where~$C_2=0$ and~$C_4=0$, and where the
boundary conditions read
\begin{equation}
\kappa_n(\lambda_p)=0,\quad\kappa_n'(\lambda_p)=0
\label{Sect:B:Eq:1}
\end{equation}
We have two conditions, but three unknowns, $C_1$, $C_3$ and~$\lambda_p$. The
last condition in~\eqref{SubSect:3.1:Eq:4}, i.e.~$\kappa''(\lambda_p)=0$, is
\textbf{not} imposed, but will be justified by direct energy minimization. 
Substituting~\eqref{SubSect:3.1:Eq:3} into~\eqref{Sect:B:Eq:1}
for~$\lambda_p>0$ yields the solution
\begin{equation}
\kappa_n(\xi)=1-\phi-\frac{1-\phi}{\cos\lambda_p\sinh\lambda_p+\cosh\lambda_p\sin\lambda_p}(\sinh\lambda_p\cos\xi+\sin\lambda_p\cosh\xi)
\label{Sect:B:Eq:2}
\end{equation}

In order to make our exposition more readable, the subsequent derivations
are structured into four steps. In~\ref{SubSect:B.1}, \new{we first normalize the functional~$\Pi^*$ from Eq.~\eqref{mj4} in order to reparametrize it in terms of the dimensionless solution~\eqref{Sect:B:Eq:2}}. Such a formulation makes it relatively easy to
demonstrate that the homogeneous interface conditions correspond to a saddle
point of the reduced energy functional, both under fixed axial force or prescribed
displacements described in~\ref{SubSect:B.2} and~\ref{SubSect:B.3}. In~\ref{SubSect:B.4} we finally demonstrate that the saddle points
are energy minima.
%
%----------------------------------
%	SUBSECTION B.1
%----------------------------------
%
\subsection{Normalization of the Reduced Functional}
\label{SubSect:B.1}
We search for the minimizers of the energy functional~$\Pi^*$, which after normalization takes the form
\begin{equation}
\overset{\circ}{\Pi}(\kappa_n)=\psi\left(\widetilde{u}-\theta\int_{-\lambda_p}^{\lambda_p}\kappa_n\,\mathrm{d}\xi\right)^2+\psi\theta\int_{-\lambda_p}^{\lambda_p}(-\kappa_n^2+\kappa_n''^2+2\kappa_n)\,\mathrm{d}\xi
\label{Sect:B:Eq:3}
\end{equation}
where we have denoted
\begin{equation}
\theta=-\frac{El}{H2l_g}
\label{Sect:B:Eq:4}
\end{equation}
In~\eqref{Sect:B:Eq:3}, $\widetilde{u}=\overline{u}/u_0$ denotes the dimensionless elongation, $u_0=2l_g\sigma_r/E=2l_g\varepsilon_0$, and~$\psi=Al_g\sigma_r^2/E$ denotes the reference energy. Since~$F=A\sigma_r\phi$, we rewrite~\eqref{mj1} as
\begin{equation}
u'-\kappa=\frac{F}{EA}=\frac{\sigma_r\phi}{E}
\label{Sect:B:Eq:5}
\end{equation}
Integration over the plastic zone~$\mathcal{I}_p$ and conversion to the normalized form provides
\begin{equation}
\phi=\widetilde{u}-\theta\int_{-\lambda_p}^{\lambda_p}\kappa_n\,\mathrm{d}\xi
\label{Sect:B:Eq:6}
\end{equation}
see also~\eqref{mj2}, which can be introduced into the energy functional~\eqref{Sect:B:Eq:3} and furnishes us with the relation
\begin{equation}
\overset{\circ}{\Pi}(\kappa_n)=\psi\phi^2+\psi\theta\int_{-\lambda_p}^{\lambda_p}(-\kappa_n^2+\kappa_n''^2+2\kappa_n)\,\mathrm{d}\xi
\label{Sect:B:Eq:7}
\end{equation}
Two situations can now be investigated: minimization under
fixed axial force~$\phi$, or under prescribed
displacement~$\widetilde{u}$. Note that, for a prismatic bar with uniform
properties in inelastic regime, we have~$\phi < 1$, which can be verified in Figs.~\ref{SubSect:3.2:Fig:3b} and~\ref{SubSect:4.1:Fig:3b} for~$\beta\rightarrow 0$ or Eq.~\eqref{SubSect:3.1:Eq:8}, and that~$\widetilde{u} > 1$ directly from its definition.
%
%----------------------------------
%	SUBSECTION B.2
%----------------------------------
%
\subsection{The case of fixed~\texorpdfstring{$\phi$}{phi}}
\label{SubSect:B.2}
Direct differentiation and integration of~\eqref{Sect:B:Eq:2} provides
\begin{subequations}
\label{SubSect:B.2:Eq:1}
\begin{align}
&\int_{-\lambda_p}^{\lambda_p}\kappa_n\,\mathrm{d}\xi=2(1-\phi)(\lambda_p-2\alpha)\label{SubSect:B.2:Eq:1a}\\
&\int_{-\lambda_p}^{\lambda_p}(-\kappa_n^2+\kappa_n''^2)\,\mathrm{d}\xi=2(1-\phi^2)(\lambda_p-2\alpha)\label{SubSect:B.2:Eq:1b}
\end{align}
\end{subequations}
where we have denoted
\begin{equation}
\alpha(\lambda_p)=\frac{\sinh\lambda_p\sin\lambda_p}{\cos\lambda_p\sinh\lambda_p+\cosh\lambda_p\sin\lambda_p}
\label{SubSect:B.2:Eq:2}
\end{equation}
Substituting from~\eqref{SubSect:B.2:Eq:1} into~\eqref{Sect:B:Eq:7} gives after some algebra
\begin{equation}
\widehat{\Pi}(\lambda_p)=\psi\left\{\phi^2+2\theta(1-\phi^2)[\lambda_p-2\alpha(\lambda_p)]\right\}
\label{SubSect:B.2:Eq:3}
\end{equation}
from which we obtain the
stationarity condition~(the prime now denotes the derivative with respect to
$\lambda_p$)
\begin{equation}
\widehat{\Pi}'(\lambda_p)=2\theta\psi(1-\phi^2)\frac{\mathrm{d}}{\mathrm{d}\lambda_p}[\lambda_p-2\alpha(\lambda_p)]=0
\label{SubSect:B.2:Eq:4}
\end{equation}
which reduces to~\eqref{SubSect:3.1:Eq:6} and hence~$\lambda_p=m_1$. Taking the second derivative provides
\begin{equation}
\left.\frac{\mathrm{d}^2}{\mathrm{d}\lambda_p^2}[\lambda_p-2\alpha(\lambda_p)]\right|_{\lambda_p=m_1}=-2\left.\frac{\mathrm{d}^2}{\mathrm{d}\lambda_p^2}\alpha(\lambda_p)\right|_{\lambda_p=m_1}=0
\label{SubSect:B.2:Eq:5}
\end{equation}
From Fig.~\ref{SubSect:B.4:Fig:1} and Eq.~\eqref{SubSect:B.2:Eq:5} we deduce that~$\lambda_p=m_1$ is a saddle point of~$\widehat{\Pi}$.
%
%----------------------------------
%	SUBSECTION B.3
%----------------------------------
%
\subsection{The case of fixed~\texorpdfstring{$\widetilde{u}$}{uD}}
\label{SubSect:B.3}
Introducing~\eqref{SubSect:B.2:Eq:1a} into Eq.~\eqref{Sect:B:Eq:6} provides
\begin{equation}
\phi(\lambda_p)=\frac{\widetilde{u}-2\theta[\lambda_p-2\alpha(\lambda_p)]}{1-2\theta[\lambda_p-2\alpha(\lambda_p)]}=1-\frac{\widetilde{u}-1}{D(\lambda_p)}
\label{SubSect:B.3:Eq:1}
\end{equation}
where we have denoted~$D(\lambda_p)=1-2\theta[\lambda_p-2\alpha(\lambda_p)]$. Substituting~\eqref{SubSect:B.3:Eq:1} into~\eqref{SubSect:B.2:Eq:3} then gives
\begin{equation}
\widetilde{\Pi}(\lambda_p)=\psi\left\{\frac{[1-D(\lambda_p)](1-2\widetilde{u})+\widetilde{u}^2}{D(\lambda_p)}\right\}
\label{SubSect:B.3:Eq:2}
\end{equation}
which is the normalized energy functional under prescribed fixed dimensionless
elongation~$\widetilde{u}$. Minimization with respect to~$\lambda_p$ yields
\begin{equation}
\widetilde{\Pi}'(\lambda_p)=-\frac{\psi(1-\widetilde{u})^2}{D^2(\lambda_p)}\frac{\mathrm{d}}{\mathrm{d}\lambda_p}D(\lambda_p)=2\psi\theta\frac{(1-\widetilde{u})^2}{D^2(\lambda_p)}\frac{\mathrm{d}}{\mathrm{d}\lambda_p}[\lambda_p-2\alpha(\lambda_p)]=0
\label{SubSect:B.3:Eq:3}
\end{equation}
The condition~$\frac{\mathrm{d}}{\mathrm{d}\lambda_p}[\lambda_p-\alpha(\lambda_p)]=0$ reduces again to Eq.~\eqref{SubSect:3.1:Eq:6}, cf also~\eqref{SubSect:B.2:Eq:4}. The second derivative of~$\widetilde{\Pi}$ provides again condition~\eqref{SubSect:B.2:Eq:5} showing that the solution is also at a saddle point.

%
%----------------------------------
%	SUBSECTION B.4
%----------------------------------
%
\subsection{Energy minima}
\label{SubSect:B.4}
\begin{figure}
\centering
\includegraphics[scale=1]{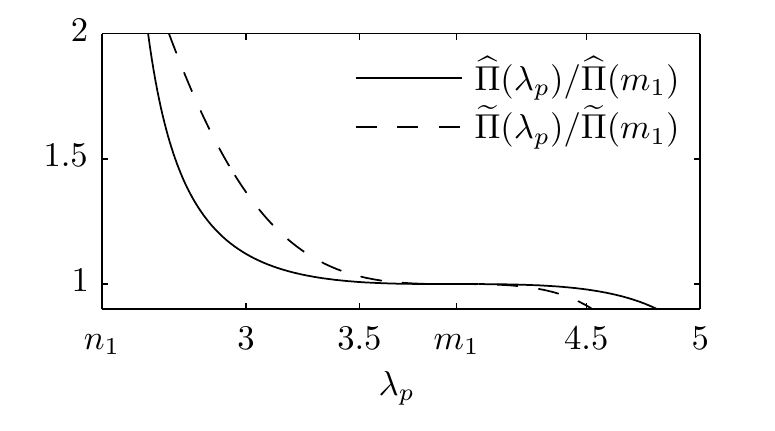}
\caption{$\widehat{\Pi}(\lambda_p)/\widehat{\Pi}(m_1)$ for fixed~$\phi=0.5$, and~$\widetilde{\Pi}(\lambda_p)/\widetilde{\Pi}(m_1)$ for fixed~$\widetilde{u}=3$ as functions of~$\lambda_p\in[n_1,m_1]$ and for a bar with uniform properties.}
\label{SubSect:B.4:Fig:1}
\end{figure}

To verify that~$\lambda_p=m_1$ is actually the minimum, we recall the geometric
constraint~$\kappa_n''(\lambda_p)\geq 0$ resulting from the requirements
that~$\kappa(\lambda_p)=0,\kappa'(\lambda_p)=0$, $\kappa(\xi)\geq 0$,
$\kappa(\xi)=0$ for~$\xi\notin(-\lambda_p,\lambda_p)$, and a Taylor series
expansion in~$\mathcal{I}_p$ near the boundary point~$\lambda_p$. Consequently,
the constraint~$\kappa_n''(\lambda_p)\geq 0$ gives for the general solution
in~\eqref{Sect:B:Eq:2}
condition~$\sinh\lambda_p\cos\lambda_p-\sin\lambda_p\cosh\lambda_p\geq 0$, which
provides the
constraint~$\lambda_p\in[n_1,m_1]$. \new{For the definition of~$m_1$ see Eq.~\eqref{SubSect:3.1:Eq:6} and the discussion below; analogously we define $n_i$ as solutions of~$\tan n_i = -\tanh n_i$, which leads to $n_{\pm i}\doteq 0,\pm 2.3650,\pm 5.4978,\dots, i\in\mathbb{N}_0$.} Since
\begin{equation}
\begin{aligned}
\widehat{\Pi}'(\lambda_p)=\widetilde{\Pi}'(\lambda_p)&=C_\Pi\frac{\mathrm{d}}{\mathrm{d}\lambda_p}[\lambda_p-2\alpha(\lambda_p)]\\
&=-C_\Pi\frac{(\cosh\lambda_p\sin\lambda_p-\cos\lambda_p\sinh\lambda_p)^2}{(\cosh\lambda_p\sin\lambda_p+\cos\lambda_p\sinh\lambda_p)^2}\ \left\{
\begin{aligned}
&<0\mbox{ for }\lambda_p\in[n_1,m_1)\\
&=0\mbox{ for }m_1
\end{aligned}
\right.
\end{aligned}
\label{SubSect:B.4:Eq:1}
\end{equation}
where~$C_\Pi>0$ is a~$\lambda_p$-independent constant, cf.
Eqs.~\eqref{SubSect:B.2:Eq:4} and~\eqref{SubSect:B.3:Eq:3}, we conclude that the
minimum is attained for~$\lambda_p=m_1$. This finding can be
visually verified in Fig.~\ref{SubSect:B.4:Fig:1}
constructed for the data from
Tab.~\ref{SubSect:3.2:Tab:1}, for~$L=2l_g=4lm_1$, and for the load parameters~$\phi=0.5$ or~$\widetilde{u}=3$.

In conclusion, for both cases, i.e.~either for fixed~$\phi$ or~$\widetilde{u}$, the boundary
condition~$\kappa_n''(\lambda_p)=0$ is indeed the optimal one and
the solution is located at a saddle point, which is at the same time
the boundary of the admissible set~$[n_1, m_1]$.
%
%-----------------------------------------------------------------------------
%	ACKNOWLEDGEMENTS
%-----------------------------------------------------------------------------
%
\section*{Acknowledgements}
Financial support of this work from the Czech Science Foundation (GA\v{C}R)
under projects No.~P201/10/0357 and 14-00420S is gratefully acknowledged.

%
%-----------------------------------------------------------------------------
%	REFERENCES
%-----------------------------------------------------------------------------
%

%\bibliographystyle{elsarticle-harv1} 
%\bibliography{mybibfile}

\end{document}